\documentclass[preprint,12pt]{elsarticle}

\usepackage{amssymb}
\usepackage{graphicx}
\usepackage{amsmath}
\usepackage{algorithm}
\usepackage{algpseudocode}
\usepackage{mathtools,nccmath}
\usepackage{booktabs}
 \usepackage{multirow}
 \usepackage{longtable}
\usepackage{makecell}
\usepackage{subfig}
\usepackage{array} 
\usepackage{hyperref}
\usepackage{booktabs,tabulary,lipsum}
\usepackage[utf8]{inputenc}
\usepackage{float}

\graphicspath{/images}

\journal{Information Fusion}

\begin{document}

\begin{frontmatter}

\title{An Advanced Data Fabric Architecture Leveraging Homomorphic Encryption and Federated Learning}

\author[inst1]{Sakib Anwar Rieyan}
\ead{sakib.anwar.rieyan@g.bracu.ac.bd}
\author[inst1]{Md. Raisul Kabir News}
\ead{md.raisul.kabir.news@g.bracu.ac.bd}
\author[inst1]{A.B.M. Muntasir Rahman}
\ead{a.b.m.muntasir.rahman@g.bracu.ac.bd}
\author[inst1]{Sadia Afrin Khan}
\ead{sadia.afrin.khan@g.bracu.ac.bd}
\author[inst1]{Sultan Tasneem Jawad Zaarif}
\ead{sultan.tasneem.jawad.zaarif@g.bracu.ac.bd}
\author[inst1]{Md. Golam Rabiul Alam}
\ead{rabiul.alam@bracu.ac.bd}
\author[2]{Mohammad Mehedi Hassan\corref{cor1}}
\ead{mmhassan@ksu.edu.sa}
\cortext[cor1]{Corresponding author}
\author[3]{Michele Ianni}
\ead{michele.ianni@unical.it}
\author[3]{Giancarlo Fortino}
\ead{giancarlo.fortino@unical.it}

\affiliation[inst1]{organization={Department of Computer Science and Engineering, School of Data and Sciences, BRAC University},
            addressline={66 Mohakhali}, 
            city={Dhaka},
            postcode={1212},
            country={Bangladesh}}

\affiliation[2]{organization={Department of Information Systems, College of Computer and Information Sciences	},
            addressline={King Saud University},
            city={Riyadh},
            postcode={11543},
           country={Saudi Arabia}
}

\affiliation[3]{organization={Department of Informatics, Modeling, Electronics, and Systems,},
            addressline={University of Calabria},
            city={Rende, CS},
            postcode={87036},
            country={Italy}}

\begin{abstract}
\textbf{Data fabric is an automated and AI-driven data fusion approach to accomplish data management unification without moving data to a centralized location for solving complex data problems. In a Federated learning architecture, the global model is trained based on the learned parameters of several local models that eliminate the necessity of moving data to a centralized repository for machine learning. This paper introduces a secure approach for medical image analysis using federated learning and partially homomorphic encryption within a distributed data fabric architecture. With this method, multiple parties can collaborate in training a machine-learning model without exchanging raw data but using the learned or fused features. The approach complies with laws and regulations such as HIPAA and GDPR, ensuring the privacy and security of the data. The study demonstrates the method's effectiveness through a case study on pituitary tumor classification, achieving a significant level of accuracy. However, the primary focus of the study is on the development and evaluation of federated learning and partially homomorphic encryption as tools for secure medical image analysis. The results highlight the potential of these techniques to be applied to other privacy-sensitive domains and contribute to the growing body of research on secure and privacy-preserving machine learning.}
\end{abstract}




\end{frontmatter}


\section{Inroduction}
\label{Inroduction}
Artificial intelligence (AI) has emerged as an indispensable component of our everyday lives, particularly within the realm of healthcare. To harness the full potential of AI in healthcare, it becomes imperative to obtain substantial quantities of meticulously curated data. Nevertheless, the preservation and examination of healthcare data encounter notable impediments due to the privacy and sensitivity associated with it. The sheer magnitude of healthcare data is frequently substantial, primarily attributable to the abundance of image-based data. Moreover, the preservation of the confidentiality of healthcare data is paramount, considering its personal and delicate nature.

To address these challenges, we have built an advanced data fabric architecture that brings together healthcare centers in a region and stores patient data and diagnoses in a secure and privacy-preserving manner. Data fabric is a data fusion \cite{34, 37} and integration approach to accomplish data management unification through analytics and AI. The proposed  approach utilizes federated learning and partially homomorphic encryption \cite{38} to allow for collaborative machine learning on encrypted data, while still maintaining compliance with laws and regulations such as the Health Insurance Portability and Accountability Act (HIPAA) \cite{1} and the General Data Protection Regulation (GDPR) Act 2018 \cite{2}.

In this study, we have used pituitary tumor classification as a case study, employing various deep-learning models such as VGG16, VGG19, ResNet50, and ResNet152. Our results show promising potential for the use of federated learning \cite{34} and partially homomorphic encryption in secure medical image analysis. Specifically, we achieved good performance with VGG16 and VGG19 models, while ResNet50 and ResNet152 achieved lower accuracy and precision for both classes. However, our custom CNN architecture outperformed all of these pre-trained models in almost every metric that we used. Our findings contribute to the growing body of research on secure and privacy-preserving \cite{35, 36} machine learning \cite{39} and demonstrate the potential for these techniques to be applied in other privacy-sensitive domains.

\subsection{Motivation}
In the field of medical image analysis, ensuring the security of sensitive patient data is of utmost importance. However, with the increasing use of machine learning and deep learning techniques for medical image analysis, there is a pressing need for an effective and secure architecture to handle such data. Previous studies have shown that the use of conventional security measures, such as encryption and access control, is not enough to ensure the privacy of patient data in the context of machine learning and deep learning operations. In addition, the use of traditional centralized architectures for processing medical image data can be slow and resource-intensive, which can further compromise the security of the data.

Therefore, there is a clear need for a new, advanced data fabric architecture that is specifically designed to handle the unique challenges of securing medical image data while also supporting efficient machine learning and deep learning operations. This research aims to address this gap in the current state of the art by proposing and evaluating a novel architecture that is capable of effectively securing medical image data while also enabling fast and accurate machine learning and deep learning operations.

\subsection{Research Problems}
The integration of data into healthcare has the potential to improve the prediction of diseases and epidemics, enhance treatment outcomes, and prevent premature deaths. However, the confidentiality of healthcare data and the complexity of managing large and diverse datasets pose significant challenges to the integration of data into healthcare. Ensuring data security and privacy is of utmost importance, as security breaches in healthcare are on the rise. According to a study \cite{3}, there were 3,033 data breaches reported between 2010 and 2019, resulting in the exposure of 255.18 million records of data.
Furthermore, the substantial volume of healthcare data presents challenges in terms of efficient processing, storage, and communication. Conventional methods may prove inadequate when confronted with the magnitude of the data at hand.
In one proposed solution \cite{4}, a big data healthcare cloud would host clinical, financial, social, physical, and psychological data from patients in a centralized location. However, proper governance of the data cloud is necessary to effectively work with and analyze complex data.

In this study, we aim to address these challenges by proposing an advanced data fabric architecture that brings together healthcare centers in a region and stores patient data and diagnoses in a secure and privacy-preserving manner using federated learning and partially homomorphic encryption. We demonstrate the effectiveness of our approach using pituitary tumor classification as a case study. However, the primary focus of our work is on the development and evaluation of federated learning and partially homomorphic encryption as tools for secure medical image analysis in the healthcare sector.

The primary objective of this study is to address the research question: 

\textit{How effective and practical is the implementation of advanced data fabric architecture using federated learning and partially homomorphic encryption for secure medical image analysis in the healthcare sector?}

\subsection{Contributions}
Through our work, we show a fully-fledged data fabric architecture based on healthcare data can be built whilst complying with privacy regulations and maintaining good accuracy scores. Our main contribution is threefold:

\begin{itemize}
  \item We propose an advanced data fabric architecture for storing and collaborating and fusing healthcare data in an encrypted form by using Partial Homomorphic Encryption (PHE)  and sharing it with other parties without revealing its content. In this architecture, medical images of various patients/clients are encrypted on the client side and these encrypted images are then used as inputs for deep learning models, enabling the models to learn and classify tumors. Subsequently, the system collects the classified tumor data for further analysis and processing. Further processing of data is done in its encrypted state. Here, The raw data was encrypted and generated to local weights using the local FL Model before getting global attention. Therefore even if the data was backtracked the end result will produce nothing but an encrypted image. Thus, this architecture provides a secure and efficient mechanism for processing encrypted data, while preserving data privacy and confidentiality regulations such as HIPAA and GDPR.
  \item The proposed federated learning framework enables multiple clients to collaboratively train machine learning models on their respective data and then store the local updates, unlike the existing general federated learning frameworks which function on real-time local and global updates. Our framework offers the flexibility to modify, scale, merge or select the local model updates before using them into the global model. In this way, the framework we proposed facilitates the systematic exchange of model updates between the local and global models.
  
  \item Moreover, we have tailored a convolutional neural network (CNN) architecture, inspired by VGG16 and VGG19, with a smaller input size, resulting in a reduced parameter size compared to the aforementioned models. This customization enables enhanced efficiency by reducing computational complexity, particularly when leveraging Partially Homomorphic Encryption (PHE) techniques.

\item We further evaluate the proposed approach by implementing a prototype of the homomorphic encryption-based data fabric and the federated learning framework. The assessment indicates that the suggested method offers an effective and reliable data fusion for sharing and analyzing data securely. The experimental results demonstrate that the proposed approach achieves satisfactory accuracy in the collaborative training of machine learning models, even when the data is encrypted.
\end{itemize}

\section{Literature Review}
\label{Literature Review}


Data fabric architecture is a relatively new concept that has already been utilized by notable organizations, including IBM, for data fusion, management, and unification purposes. Despite its potential benefits, there is limited research available on the implementation of this architecture in the healthcare system. Due to the sensitive nature of healthcare, developing a secure data fabric architecture can present challenges. This chapter examines various data architectures, processes, and encryption methods that can be employed to ensure the security of healthcare data.

There are a few works that are related to the architecture we are working on. They are described below:
 
In \cite{9}, the authors describe a proof-of-concept implementation that uses the Hyperledger Fabric framework. They claim that this concept is capable of storing patient records effectively with keeping all the privacy protocols intact. Lastly, they compare the read-write times according to their claim.

In \cite{10}, Roehrs et al. divide personal health record (PHR) information into data blocks that may appear to be centrally stored but are actually distributed among participating devices. The authors claim that their proposed openPHR protocol is practical, flexible, and scalable for adoption by multiple organizations. Although the authors provide a detailed architecture, questions have been raised about the feasibility of their approach, especially regarding security and privacy concerns. It is important to note that PHRs are controlled by patients, while electronic health records (EHRs) are managed by healthcare institutions. Nevertheless, EHRs and PHRs are electronically stored and distributed and can be assessed based on metrics such as performance, scalability, privacy protection, and compliance with the GDPR.

In \cite{11}, the MeDShare platform shares several similarities with PREHEALTH \cite{9}. However, the authors do not explicitly specify the underlying blockchain framework. Additionally, their emphasis is more on examining the fundamental components of blockchain technology, including data blocks and smart contracts, rather than presenting a practical solution.

Ming and Zhang \cite{12} present an effective privacy-preserving access control (PPAC) strategy for cloud-based EHR systems. Their approach utilizes the cuckoo filter and an innovative attribute-based signcryption (ABSC) mechanism to achieve both anonymity and computational efficiency. The authors offer comprehensive assurances of privacy and conduct thorough performance evaluations for comparison. However, it is uncertain whether their approach complies with the GDPR regulations.

In \cite{13}, Fu et al. mainly focus on sharing data among different participants safely. Using Hyperledger Fabric, they propose a more secure decentralized distributed data storage over traditional centralized data for SKA data due to high management costs and low credible traceability. The SKA Data Management Alliance significantly reduces costs and improves the overall security of its data by adopting this distributed storage system.

Our architecture utilizes federated learning to train deep learning models. In our research, we have come across some related studies that are somewhat aligned with our work. Here are brief descriptions of these studies:

In \cite{16}, the authors propose a secured model of federated learning using homomorphic encryption and attempt to classify Covid-19 using X-Ray images. They secure the federated process and claim that sensitive information can fall into the hands of attackers if the DL process is not secured. They do not encrypt the dataset but rather encrypt the weight matrix used in federated learning. They claim to achieve an accuracy of 84.00\% with a precision of 86.89\%.

It is obvious that with the vast growth of artificial intelligence and big data, contradictions between user data policy and data policy also grow proportionally. That is why in \cite{18}, the authors come up with a vertical federated learning system for Bayesian machine learning using homomorphic encryption. Their model can be compared with 90\% of models trained by a single union server. Additionally, their system can be used in education, finance, medicine, risk controls, and other fields.

Here are some additional studies that are relevant to our research and have been reviewed and summarized below:

In \cite{19}, the authors provide insights into the difficulties of medical data analysis and security and propose a solution based on a decentralized architecture. They utilize the Exonum framework. In their proposed architecture, they separate the whole system into two parts - 'Closed Information' and 'Open Information.' In the closed part, encrypted data is stored in a blockchain, while in the open part, non-encrypted service-related data is stored.

Zhang et al. \cite{20} propose a meaningful usage of optimizing Electronic Health Records (EHRs) using big data analytics. Here, they propose an insight into how to improve electronic health records using three methods: Data Collection, Data Storage, and Data Utilization. Firstly, in the data collection method, records are divided into structured and unstructured data. Structured data includes demographics, health status, lab results, billing, etc., while surgical videos or diagnosis notes fall under unstructured data. After collecting, they propose a transformation engine where data is moved, cleaned, merged, and validated, and is stored in DBMS, Cloud, or NoSQL. Finally, transformed data is processed using mapping and reduction, and stream computing and in-database analytics are used for generating reporting systems, which help achieve a meaningful usage of EHRs using big data analytics. However, the authors also map out the limitations of this research, and emphasize that it lays the foundation for interesting opportunities in the future.

In \cite{23}, the authors introduce a novel big data platform that can redesign modern medical data and bring an effective and quick solution to the healthcare system. They propose a system that is lightning-fast, supports stream processing, and integrates with both NoSQL and RDBMS. This process aims to exploit open-source technologies as much as possible and build the system on top of them. The core of the system is Spark core, and to utilize it, the system uses the Spark framework with real-time graphical image processing. For handling structured data, the authors propose SparkSQL Structured Data and MLib Machine Learning for classifying the data.

In \cite{24}, the authors describe an architecture where an improvised big data model is involved to create a cloud computing environment for healthcare. In this process, huge amounts of data from medical sources and processes are collected in cloud storage, and real-time analysis is done using cloud computing for better accuracy. The Healthcare Data Management Framework is built on Hadoop Clusters and certain key components. Semantic Practitioner, Big data container and processing layer, Query formulator, Batch scheduling, and Data reader are examples of such tools. This architecture is also open to implementing various cryptographic techniques on the Cloud.

\subsection{Background Studies}
\label{Background Studies}

\subsection{Data Fabric}
According to Gartner \cite{5}, Data Fabric is a unified and integrated platform that enables data discovery, fusion and integration, management, and access across multiple environments. It provides a consistent and scalable approach to managing data assets that are distributed across various locations, such as on-premises, cloud, and edge computing. Data Fabric helps organizations to simplify and optimize their data management processes, reduce data silos, and enable real-time access to data. It also supports the creation of a self-service data marketplace, allowing users to discover, share, and consume data in a secure and governed manner. Data Fabric is increasingly becoming a critical component of modern data architectures, as organizations seek to manage the growing volume, velocity, and variety of data generated by digital business initiatives.

In our research, we are utilizing homomorphic encryption to classify pituitary tumors from MRI images in our dataset. We have used a Data Fabric architecture to store the weights of different machine-learning models as encrypted data. The ML models are run on client PCs, and the resulting encrypted data is saved in our Data Lake. Using homomorphic encryption, a server PC can perform computations on the encrypted data, allowing for the creation of a homogeneous global model. The server can then provide users with the requested results without compromising the privacy of the MRI images. This approach benefits from the Data Fabric's ability to provide a unified and integrated platform that enables data discovery, integration, management, and access across multiple environments. By utilizing homomorphic encryption and a Data Fabric architecture, we can classify pituitary tumors from MRI images in a privacy-preserving manner, contributing to the development of more secure and privacy-preserving medical imaging technologies.

\vspace{0.3cm}
\subsubsection{Vanilla Architecture of Data Fabric}

\begin{enumerate}[i)]
  \item \textbf{Accessing Data:} 
        \begin{enumerate}
            \item \textbf{Data Collecting and Encryption:} Data is a volatile resource, and Medical Data is considered highly sensitive as it can contain personally identifiable information such as names, addresses, dates of birth, and medical records which can be exploited if they fell into wrong hands. To comply with this shortcoming, in this architecture, data is neither collected nor stored in a central server which may possess the risk of data leakage.

            Here, firstly, to ensure privacy and reduce data volatility, medical data from various users are first selected and then encrypted with Partially Homomorphic Encryption (PHE). Subsequently, the encrypted data is selected to train the model locally for collecting updated model weights. The data of each user is generated and stored locally, without being transferred to the central server. Instead, the generated model updates are stored and merged for the global model formation.
            
            \item \textbf{Master Data Management:} 
            Following the generation of local model updates and subsequent merging of data, feature selection is employed as a means of optimizing and enhancing efficiency. By selecting the most relevant features or weights, the dimensionality of the data can be reduced, facilitating ease of analysis. Moreover, feature selection mitigates the risk of overfitting, improves model accuracy, and reduces computational costs, thereby achieving heightened efficiency through the utilization of a reduced training dataset.
            
            FedMax, FedAvg, and FedMin are optimization algorithms used in Federated Learning for feature selection. In all three algorithms, updated model weights are sent to the server/stored for future usage. However, in the case of FedMax, the server/user selects the model with the highest accuracy while it chooses the lowest loss model for FedMin and an average of all models for FedAvg.

            In our architecture, we selected FedMax as our feature selection algorithm to select important and relevant model weights as it showed more accuracy and efficiency compared to FedAvg and FedMin. The selected data is collected and kept together as “Master Data”.
        \end{enumerate}

    \item \textbf{Managing Life Cycle:} 
        \begin{enumerate}
            \item \textbf{Governance:} Data governance is an essential component of data fabric architecture. Data fabric architecture is an approach to data management that enables organizations to manage and process data from multiple sources, locations, and formats. It provides a unified view of data across the organization and supports various data processing requirements, such as data integration, analytics, and artificial intelligence.

            Data governance in data fabric architecture refers to the policies, processes, and standards that organizations implement to manage their data assets effectively. Data governance helps organizations ensure that their data is accurate, consistent, and compliant with regulatory requirements. It also helps organizations manage data privacy, security, and access.

            The following are some key considerations for data governance in data fabric architecture:

            \textbf{Data quality:} Data governance policies should include measures to ensure data quality, such as data profiling, data cleansing, and data validation.

            \textbf{Metadata management:} Data governance policies should include metadata management to ensure that data is properly tagged, categorized, and classified. Metadata helps organizations understand the meaning and context of their data and facilitates data discovery and reuse.

            \textbf{Data privacy and security:} Data governance policies should include measures to ensure data privacy and security, such as access controls, data encryption, and data masking.

            \textbf{Data lineage:} Data governance policies should include data lineage to track the origin, transformation, and movement of data across the organization. Data lineage helps organizations understand how data is used and facilitates compliance with regulatory requirements.

            \textbf{Data ownership and stewardship:} Data governance policies should define data ownership and stewardship to ensure that data is managed and maintained by the appropriate individuals and teams.

            \item \textbf{Compliance: } Data compliance refers to adhering to relevant laws, regulations, and industry standards related to the handling, processing, and storage of data. In the context of data fabric, data compliance refers to ensuring that data is managed in accordance with these requirements across the entire data fabric. To ensure data compliance in a data fabric, it is necessary to establish policies and procedures that cover the entire data lifecycle, from data ingestion to archival and deletion. Personal data must be collected, processed, and stored in compliance with privacy regulations such as GDPR, CCPA, HIPAA, etc.
            
            In this architecture, the feature selected weights were trained on various models such as VGG16, VGG19, ResNet 50, ResNet 152, etc. and updates are stored in a data lake structurally based on models they were trained on complying with HIPAA regulations.

        \end{enumerate}

    \item \textbf{Exposing Data:} Data exposure refers to making data available for consumption and analysis by users or applications within an organization. Exposing data in a data fabric involves providing access to the data for authorized users or applications. There are several ways to expose data in a data fabric, including:
    
    \textbf{APIs:} Application Programming Interfaces (APIs) enable applications to access and retrieve data from the data fabric.

    \textbf{Data Catalogs:} A data catalog provides a searchable inventory of data assets in the data fabric. Users can discover and access data assets through the data catalog.

    \textbf{Self-Service Analytics:} A self-service analytics platform enables users to create their own queries and reports using the data available in the data fabric.

    \textbf{Data Virtualization:} Data virtualization enables users to access and combine data from multiple sources as if it were in a single location.
  
\end{enumerate}

\begin{figure}[htbp]
 \centering
  \includegraphics[scale=0.25]{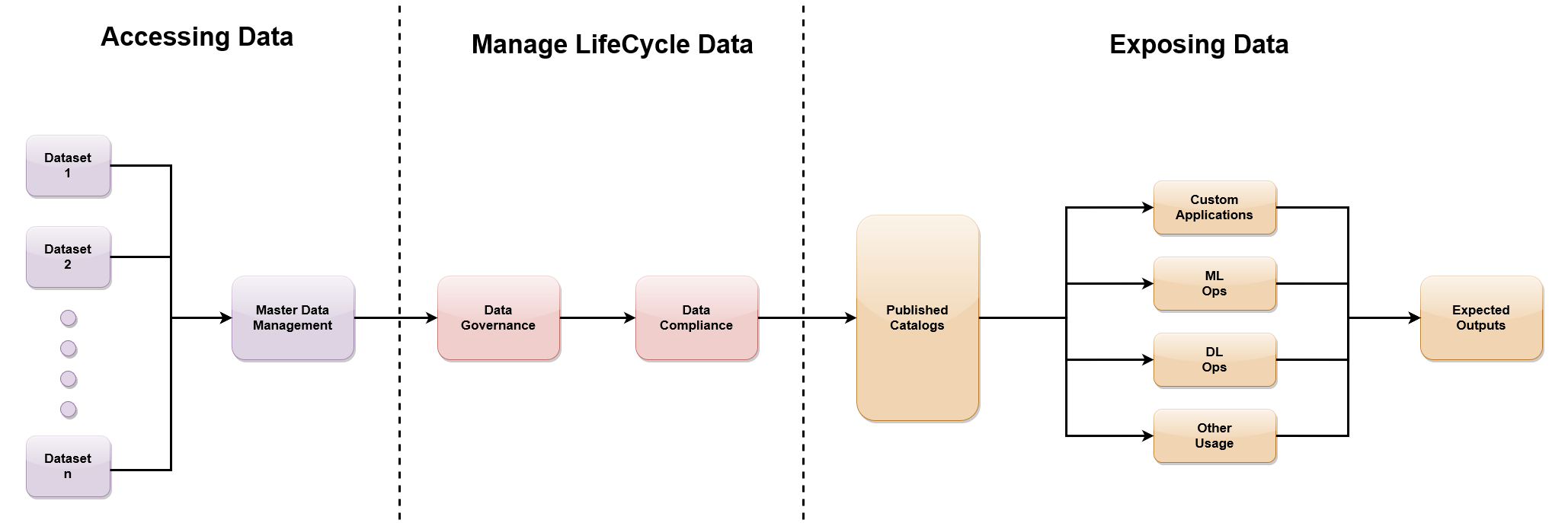}
\caption{Vanilla Architecture of Data Fabric}
\label{fig:x Vanilla Architecture of Data Fabric}
\end{figure}

\subsection{Federated Learning}
Federated Learning is a distributed machine learning technique that enables multiple clients to collaboratively learn a shared model without exchanging their raw data. This technique has gained popularity in recent years due to its ability to preserve data privacy and security while improving model performance. In Federated Learning, each client trains a local model using its own data and then sends the local model weights to a central server. The central server then aggregates the local model weights to update a global model that is shared among all clients. This process continues iteratively until the global model achieves the desired level of accuracy. According to the report \cite{6}, Federated Learning has been successfully applied to various domains, such as speech recognition, natural language processing, and healthcare, where data privacy is a major concern.


\begin{algorithm}
	\caption{Federated Learning Process}
	\begin{algorithmic}[1]
        \State Initialize global model
        \State Split data into shards and distribute among clients 
		\For{ i in range(num\_communication\_rounds) }
		\State Initialize local model
        \State Send local model to each client 
		\For{ client in clients\_data }
        \State client.model.train(client.data) 
        \EndFor
		\State Clients send updates to the global model 
        \State Scale updates by the weight scaling factor 
        \State Aggregate updates using FedMax 
        \State \vspace{-0.82cm}
          \begin{fleqn}[\dimexpr\leftmargini-\labelsep]
          \setlength\belowdisplayskip{0pt}
          \begin{equation*}
              \begin{multlined}
                global\_model.update\_weights(max(updates) \\
                \times weight\_scaling\_factors)
              \end{multlined}
          \end{equation*}
          \end{fleqn}
        \EndFor
    \end{algorithmic}
\end{algorithm}

\newpage

 \begin{figure}[htbp]
  \centering
  \includegraphics[scale=0.4]{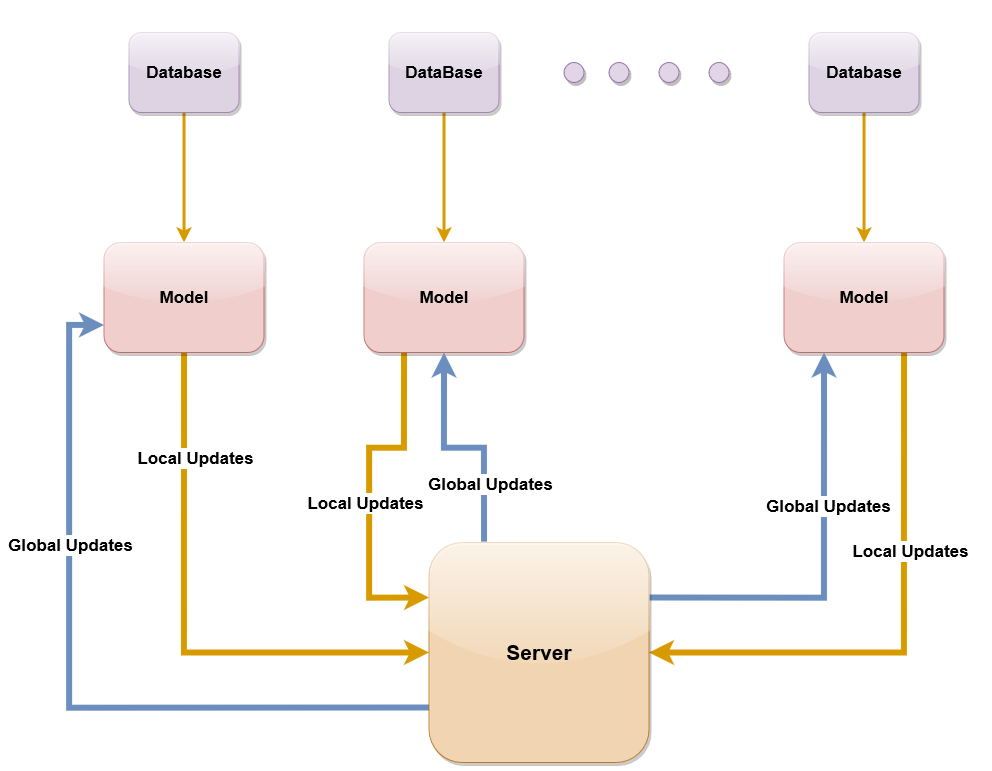}
  \caption{Federated Learning Model \cite{33}}
  \label{fig:x Federated Learning Model}
  \end{figure}

\subsection{Homomorphic Encryption}
A cryptographic method called homomorphic encryption enables mathematical operations to be carried out on ciphertext without exposing the underlying plaintext. In our research, we used partially homomorphic encryption to encrypt sensitive medical images, specifically brain MRI scans.

Partially homomorphic encryption (PHE) is a type of homomorphic encryption that only supports a limited set of mathematical operations, such as addition or multiplication. By encrypting the medical images using this technique, we were able to process and analyze the data without exposing the sensitive information contained within it.

One major benefit of using partially homomorphic encryption in this context is that it ensures the confidentiality of medical data. As medical information is often highly sensitive and personal, it is important to protect it from unauthorized access. By encrypting the data, we were able to securely process and analyze it without compromising its confidentiality.

In addition, partially homomorphic encryption allows for more efficient processing of the encrypted data. Because the mathematical operations can be performed directly on the ciphertext, there is no need to decrypt the data first, which can be a time-consuming process. This was particularly useful when working with large datasets or when processing data in real time. \cite{7}

Overall, our use of partially homomorphic encryption proved to be a successful and effective method for protecting the confidentiality of sensitive medical images while still enabling their analysis.

\renewcommand{\arraystretch}{1.5}
\begin{table}[htbp]
\centering
\caption{Comparison among different types of Homomorphic Encryption}
\resizebox{\columnwidth}{!}{%
\begin{tabular}{|cccc|}
\hline
\multicolumn{1}{|c|}{\textit{\textbf{\begin{tabular}[c]{@{}c@{}}Types of Homomorphic\\ Encryption\end{tabular}}}} & \multicolumn{1}{c|}{\textit{\textbf{\begin{tabular}[c]{@{}c@{}}Partially Homomorphic\\ Encryption (PHE)\end{tabular}}}} & \multicolumn{1}{c|}{\textit{\textbf{\begin{tabular}[c]{@{}c@{}}Somewhat Homomorphic\\ Encryption (SHE)\end{tabular}}}} & \textit{\textbf{\begin{tabular}[c]{@{}c@{}}Fully Homomorphic\\ Encryption (FHE)\end{tabular}}} \\ \hline
\multicolumn{4}{|l|}{} \\ \hline
\multicolumn{1}{|c|}{\textbf{Supported Operations}} & \multicolumn{1}{c|}{Addition or Multiplication} & \multicolumn{1}{c|}{Addition \& Multiplication} & Arbitrary Computations \\ \hline
\multicolumn{1}{|c|}{\textbf{Security}} & \multicolumn{1}{c|}{High} & \multicolumn{1}{c|}{Moderate to High} & High \\ \hline
\multicolumn{1}{|c|}{\textbf{Computational Efficiency}} & \multicolumn{1}{c|}{High} & \multicolumn{1}{c|}{Moderate} & Low \\ \hline
\multicolumn{1}{|c|}{\textbf{Computational Intensity}} & \multicolumn{1}{c|}{Low} & \multicolumn{1}{c|}{Moderate} & High \\ \hline
\multicolumn{1}{|c|}{\textbf{Encryption}} & \multicolumn{1}{c|}{Simple} & \multicolumn{1}{c|}{More Complex than PHE} & Very Complex \\ \hline
\multicolumn{1}{|c|}{\textbf{Encryption Overhead}} & \multicolumn{1}{c|}{Low} & \multicolumn{1}{c|}{Moderate} & High \\ \hline
\multicolumn{1}{|c|}{\textbf{Implementation Ability}} & \multicolumn{1}{c|}{Easy} & \multicolumn{1}{c|}{Moderate} & Difficult \\ \hline
\end{tabular}%
}

\label{table:1}
\end{table}

\begin{figure}[htbp]
  \centering
  \includegraphics[scale=0.43]{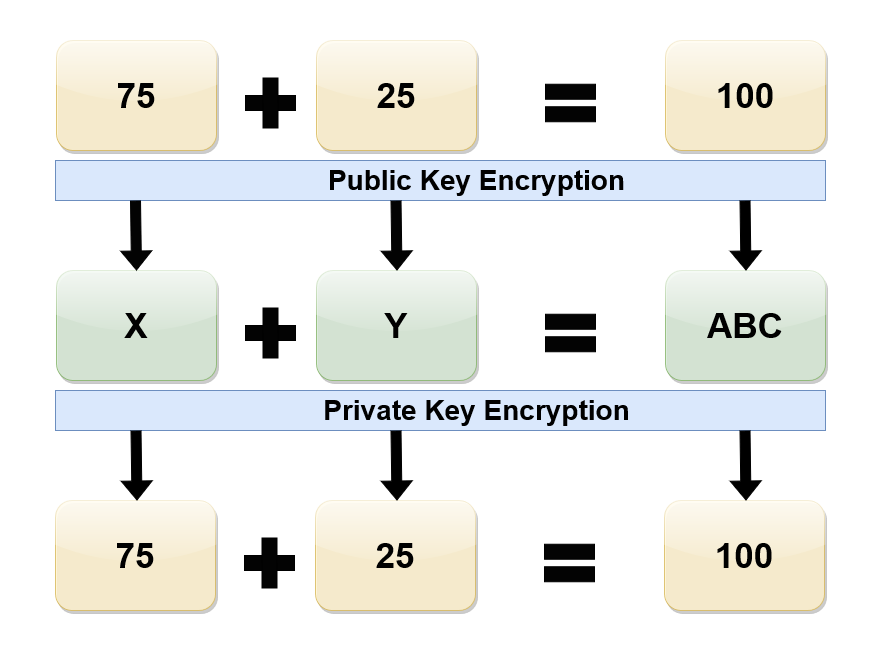}
  \caption{An Overview of the Homomorphic Encryption}
  \label{fig:x Homomorphic Encryption}
  \end{figure}

\section{Methodology}
\subsection{Data Description} 
In our experimental evaluation, we used
the dataset \cite{29} named ``Brain MRI Images for Brain Tumor Detection'', available on Kaggle\footnote{\url{https://www.kaggle.com/datasets/masoudnickparvar/brain-tumor-mri-dataset}}. The dataset includes a large number of 2D MRI images for the classification of brain tumors. Brain tumors are classified into three types: Benign, Malignant, and Pituitary. However, for our selected dataset, images are classified into two categories; Pituitary Tumor and No Tumor. 

From the dataset, 1852 images have been used by us as shown in the figure below. On the images, we performed encryption and machine learning techniques for ensuring privacy and achieving desired results. 

\begin{figure}[htbp]
\centering
\includegraphics[scale=0.45]{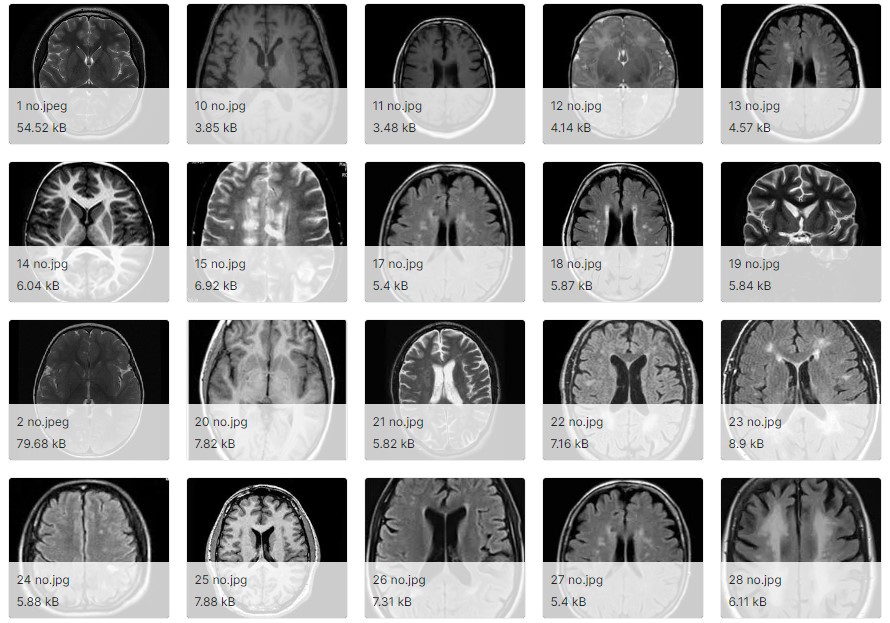}
\caption{A few samples of the used Brain Tumor dataset \cite{29}}
\label{fig:x Collected dataset from Kaggle}
\end{figure}

For preparing our own data fabric to perform MLOps, an image was selected from the dataset and converted into a Numpy array with its pixel values. After that, homomorphic encryption was performed on the array to make the data fabric encrypted. A brief overview is shown in the following figure.

\begin{figure}[htbp]
\centering
\includegraphics[width=\linewidth]{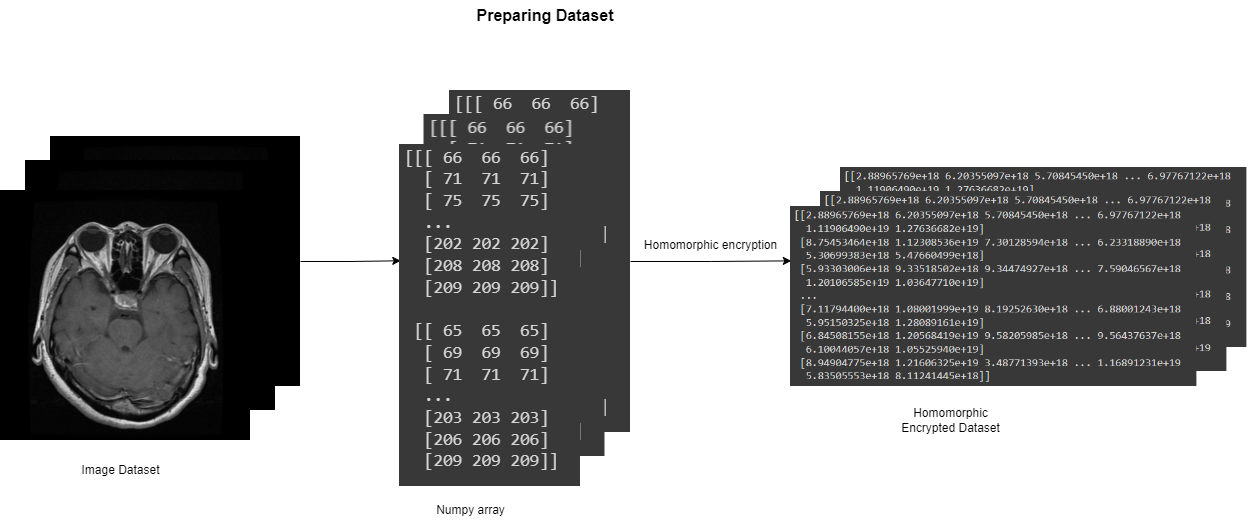}
\caption{Overview of the encrypted dataset}
\label{fig:x Overview of the encrypted dataset}
\end{figure}


\subsection{Data Preprocessing}
To prepare our data we needed to preprocess the whole dataset. Firstly, we have encrypted our dataset using the Partially Homomorphic Encryption Algorithm. After that, we resized the images to $128 \times 128$ dimensions. The shape of our dataset became (1852, 128, 128). Then we reshaped our dataset by multiplying the dimensions of individual pixels. Then the shape ultimately (1852, 15376). For machine learning classifiers we have scaled the image dataset to value 0 to 1.

\subsection{Proposed Model}
\subsubsection{Advanced Architecture of Data Fabric}

\vspace{0.1cm}
\begin{enumerate}[i)]
  \item \textbf{Accessing Data:} Here, initially, to ensure privacy and reduce data volatility, medical data from various users are first selected and then encrypted with Partially Homomorphic Encryption (PHE). Partially Homomorphic Encryption (PHE) enabled us to perform DLOps on the data securely. Subsequently, the encrypted data is selected to train the model locally for collecting updated model weights. For training, various federated learning models like VGG16, VGG19, ResNet50, ResNet152, and our Custom CNN were used which generated model updates for each model. The data of each user is generated and stored locally, without being transferred to the central server. Instead, the local model updates were stored and merged for the global model formation. 

 After combining the local model weights, we selected FedMax as our feature selection algorithm to select important and relevant model weights as it showed more accuracy and efficiency compared to FedAvg and FedMin. The selected data is collected and kept together as “Master Data”, with which we will continue our next works. 

\begin{figure}[htbp]
  \centering
  \hspace*{0.4cm}
  \includegraphics[width=\columnwidth]{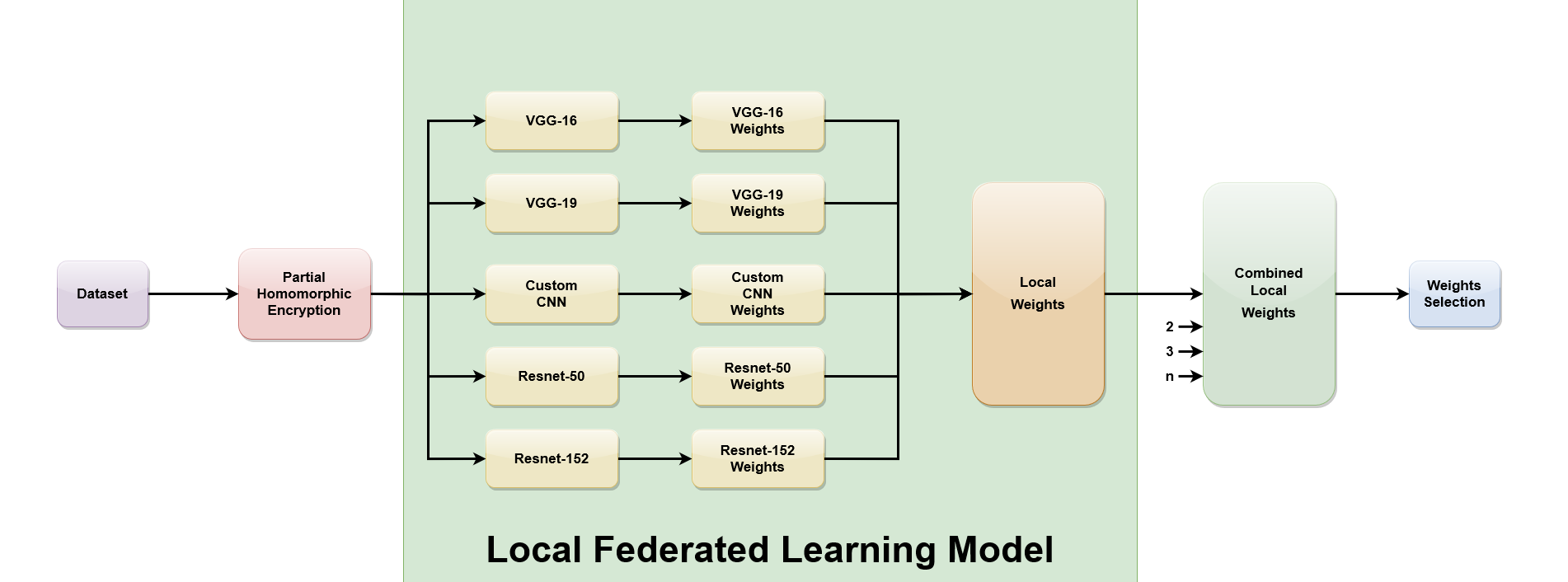}
  \caption{Accessing Data}
  \label{fig:x Accessing Data}
  \end{figure}
 
 \item \textbf{Managing Life Cycle:} Selected model weights were differentiated based on the models they were trained on. In this case, for example, updates of the VGG16 model were kept structurally under the “VGG16” name. This enables effective data governance, as the risks of data inconsistency and complicated integration across the whole architecture get lowered. Additionally, organizing data in a structured way ensures proper usage of data and helps strike a balance between data collaboration and privacy mandates. 
            
Furthermore, as the privacy of the data was already ensured in the first step, the collected model weights already comply with existing privacy regulations such as HIPAA, GDPR, etc. 

\begin{figure}[htbp]
  \centering
  \captionsetup{justification=centering}
  \hspace*{0.7cm}
  \includegraphics[scale=0.7]{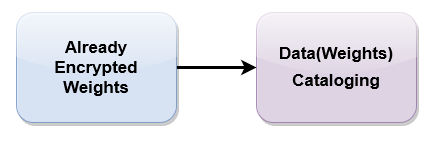}
  \caption{Data Lifecycle Management}
  \label{fig:x Data Lifecycle Management}
  \end{figure}

\item \textbf{Exposing Data:} Since exposing data is a comprehensive approach where appropriate data access controls, data fusion, and cataloging must be implemented, we stored all the collected local model updates in a “Data Lake”. This enabled us to store our raw model weights structurally in their native format. Additionally, since there were huge amounts of model weights, using a Data Lake helped us to store and process it easily. 
    
Most importantly, we can also perform MLOps on the Data Lake directly. On the client side, clients can train their data on the global federated model and can compare the global weights with the local weights of that model stored in the data lake and generate the desired output. 

\begin{figure}[htbp]
  \centering
  \includegraphics[scale=0.5]{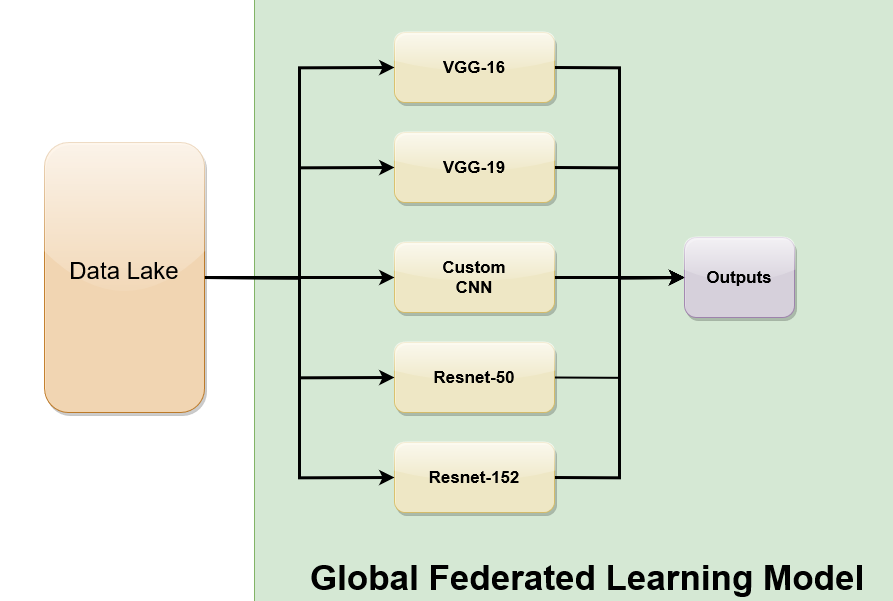}
  \caption{Exposing Data}
  \label{fig:x Exposing Data}
  \end{figure}

\end{enumerate}

\newpage
  \begin{figure}[htbp]
  \centering
  \includegraphics[width=\linewidth]{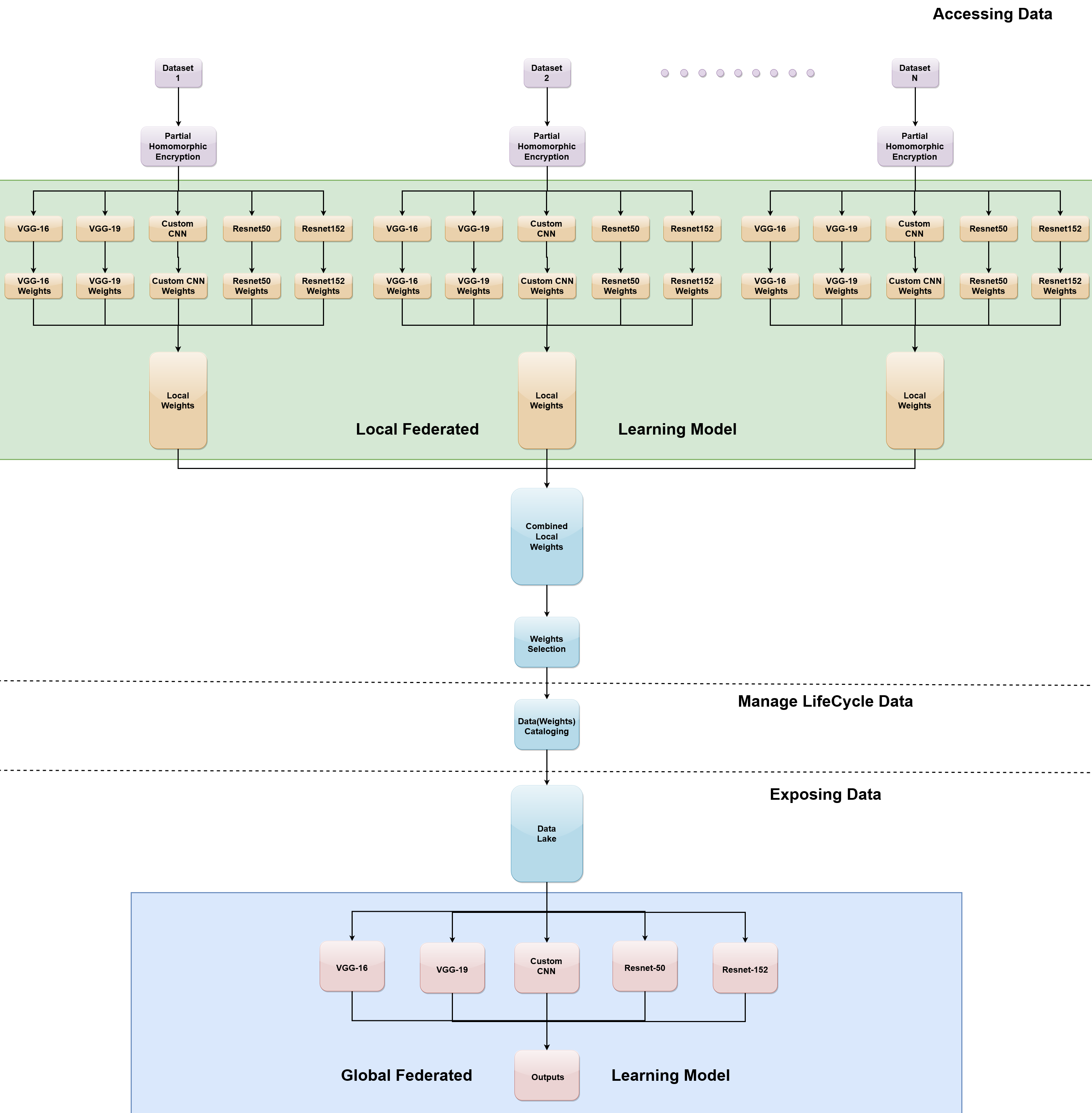}
  \caption{The Proposed Data Fabric Architecture leveraging Federated Learning and Homomorphic Encryption}
  \label{fig:x Proposed Architecture of Data Fabric}
  \end{figure}

\subsubsection{Implemented Federated Learning Framework}


As we delve into the world of medical imaging and machine learning, we have utilized a cutting-edge approach to store the local weights of our machine learning models using a Data Fabric architecture. This architecture has allowed us to securely store and manage distributed data across multiple environments, providing a consistent and scalable approach to managing data assets. 

\begin{figure}[htbp]
  \centering
  \includegraphics[scale=0.42]{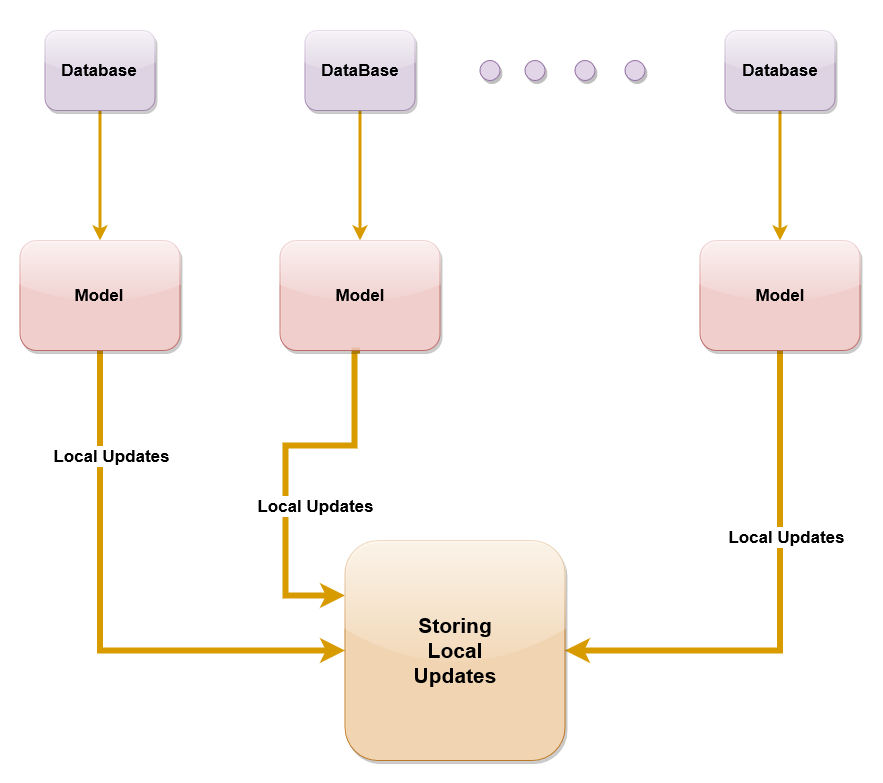}
  \caption{Local Federated Learning Model}
  \label{fig:x Local Federated Learning Model}
  \end{figure}

Our goal was to develop a pituitary tumor classification model that leverages the decentralized data in a privacy-preserving manner, improving the accuracy of the model while ensuring the privacy of patient information. We have used the FedMax algorithm, an improvement on the popular FedAvg algorithm, to compile a global model and local models for the classification of pituitary tumors from MRI images in our dataset. The FedMax algorithm considers the model performance of each client and assigns a weighting factor that allows clients with the best performance to contribute more to the global model. We have stored the resulting model weights in a data lake, allowing us to generate a global model and test it using test cases when the user prompts to show results.

\newpage
\begin{figure}[htbp]
\centering
\includegraphics[scale=0.42]{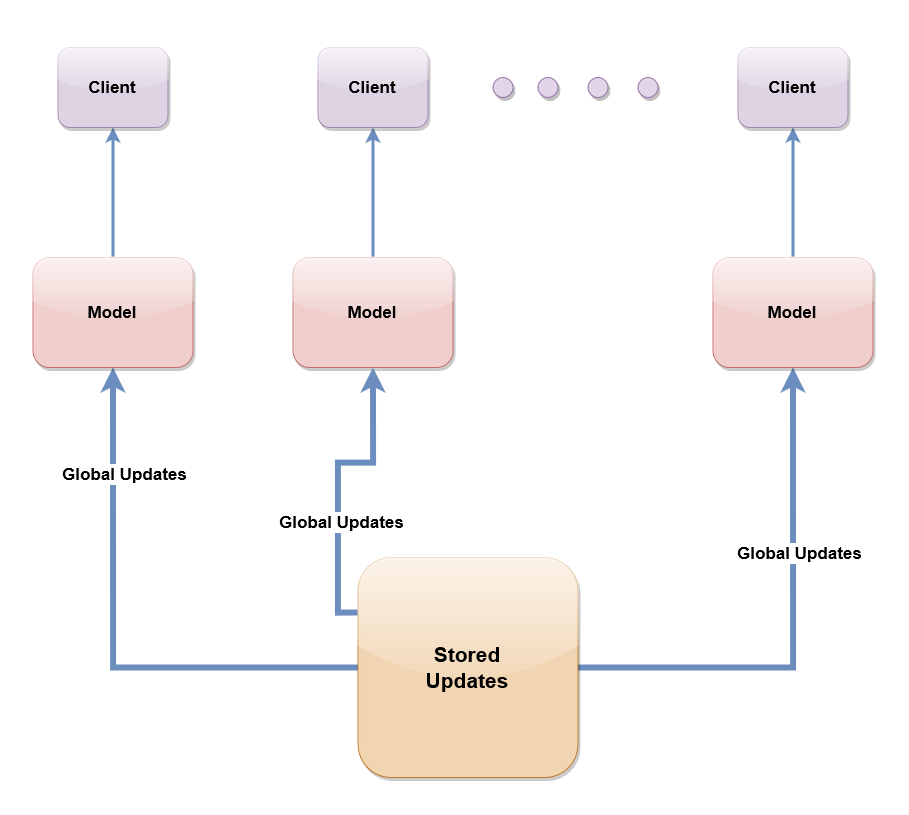}
\caption{Global Federated Learning Model}
\label{fig:x Global Federated Learning Model}
\end{figure}

\subsubsection{Comparative Analysis with Related Works}

Our work is different from existing approaches as it enables effective and secure handling of highly sensitive health data through our proposed Data Fabric architecture. Another advantage of this work is that its privacy-preserving features (Homormophic encryption) are compliant with GDPR \cite{30}, as it supports the right to be forgotten, as the actual health data is neither collected nor stored. Rather, after training, the encrypted data which is collected is stored as model updates. Similarly, it is possible to delete a client’s data from the data lake upon request. In addition to this, compared to other works, our model also complies with HIPAA guidelines, as it ensures confidentiality, integrity, and availability of personal health information, safeguards data from threats, and protects impermissible access as mentioned here\cite{1}.

The following table displays a comparison between our data fabric architecture and other existing works that have demonstrated proof-of-concept implementations, which can be implemented and practical for real-world scenarios. In this table, the technology, privacy-preserving features, GDPR and HIPAA compliance, and also performance assessment results of each existing proposal have been provided.

\renewcommand{\arraystretch}{1.5}

\begin{table}[htbp]
\centering
\caption{Comparative Analysis}
\resizebox{\columnwidth}{!}{%
\begin{tabular}{|c|c|c|c|c|c|c|}
\hline
Architecture & Technology & Access Level & HIPAA & GDPR & \begin{tabular}[c]{@{}c@{}}Privacy \\ Preserving\end{tabular} & Performance \\ \hline
PREHEALTH \cite{9} & Hyperledger Fabric & Private & Not Mentioned & Yes & Yes & Yes \\ \hline
OmniPHR \cite{10} & Peer - to - Peer & Private & No & No & No & Yes \\ \hline
MeDShare \cite{11} & Agnostic & Open & No & No & No & Yes \\ \hline
\begin{tabular}[c]{@{}c@{}}Access Control\\Based EHR \cite{12}\end{tabular} & AC Scheme & Private & No & No & Yes & Yes \\ \hline
Our Proposed Work & Data Fabric & Private & Yes & Yes & Yes & Yes \\ \hline
\end{tabular}%
}

\label{table:2}
\end{table}

\subsubsection{Model Specification}

\vspace{0.2cm}
\begin{enumerate}[i)]
  \item \textbf{VGG16:} A 16-layered convolutional neural network model trained on the ImageNet dataset, it is considered to be one of the best models to date. It is widely regarded for its simple architecture and excellent image classification performance, retaining 92.7\% test accuracy in the ImageNet dataset, which consists of almost 14 million training images across a thousand object classes.
  
  As its name suggests, it is composed of 16 layers which include 13 convolutional layers and 3 fully connected layers. It comprises 138 million parameters  and uses small convolutional filters and deep architectures to gain a large receptive field and strong discrimination ability, which helps in image classification, object detection, and semantic segmentation. To control overfitting and reduce spatial dimensions, its architecture has five max pooling layers.
  
  The most unique thing about VGG16 is that it is focused on convolution layers of $3 \times 3$ filter with stride 1 and always used the same padding and max pol layer of $2 \times 2$ filter with stride 2, and the layers are constantly arranged over the whole architecture. Due to its high-level feature representation, it provides good performance on object detection, fine-grained image classification, etc. \cite{32}

  \newpage
  \begin{figure}[htbp]
  \centering
  \hspace*{0.4cm}
  \includegraphics[width=\columnwidth]{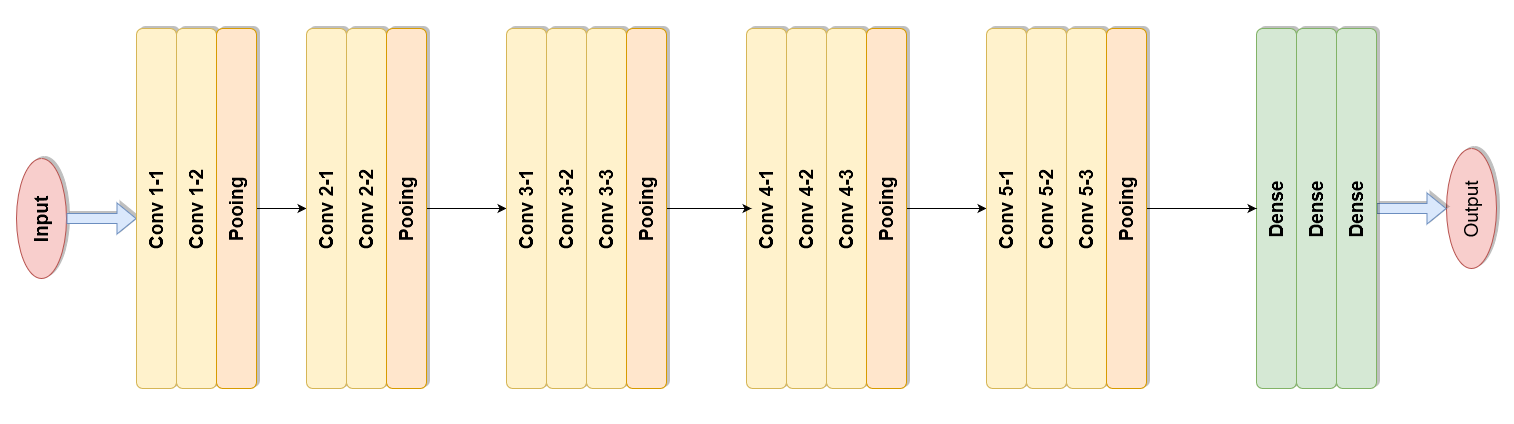}
  \caption{VGG16 Model Architecture}
  \label{fig:x VGG16 Model Architecture}
  \end{figure}

  \item \textbf{VGG19:} A variant of the VGG neural network, it is a 19-layer version of the VGG network and similar to the VGG16 architecture. Compared to VGG16, it has 5 convolutional layers and 1 fully connected layer. It has around 143 million parameters and is trained on a dataset with 1.2 million images and 1000 classes.
  
  The function accepts an image of shape (128, 128, 1) as input, and the image is passed through concatenate layer which concatenates the image 3 times resulting in a tensor with shape (128, 128, 3), this is passed to a VGG19 model which is a pre-trained convolutional neural network model that is trained on the ImageNet dataset. The VGG19 model serves as a feature extractor and it extracts features from the image by stacking convolutional layers and pooling layers. The output of the final max pooling layer is then passed through a flattened layer which reshapes the output tensor into a 2D array. This flattened layer's output is passed through a dense layer with 1 unit and a sigmoid activation function, it produces the final output of the network which represents the predicted probability of the input image belonging to the target class. \cite{32}

  \newpage
  \begin{figure}[htbp]
  \centering
  \includegraphics[width=\columnwidth]{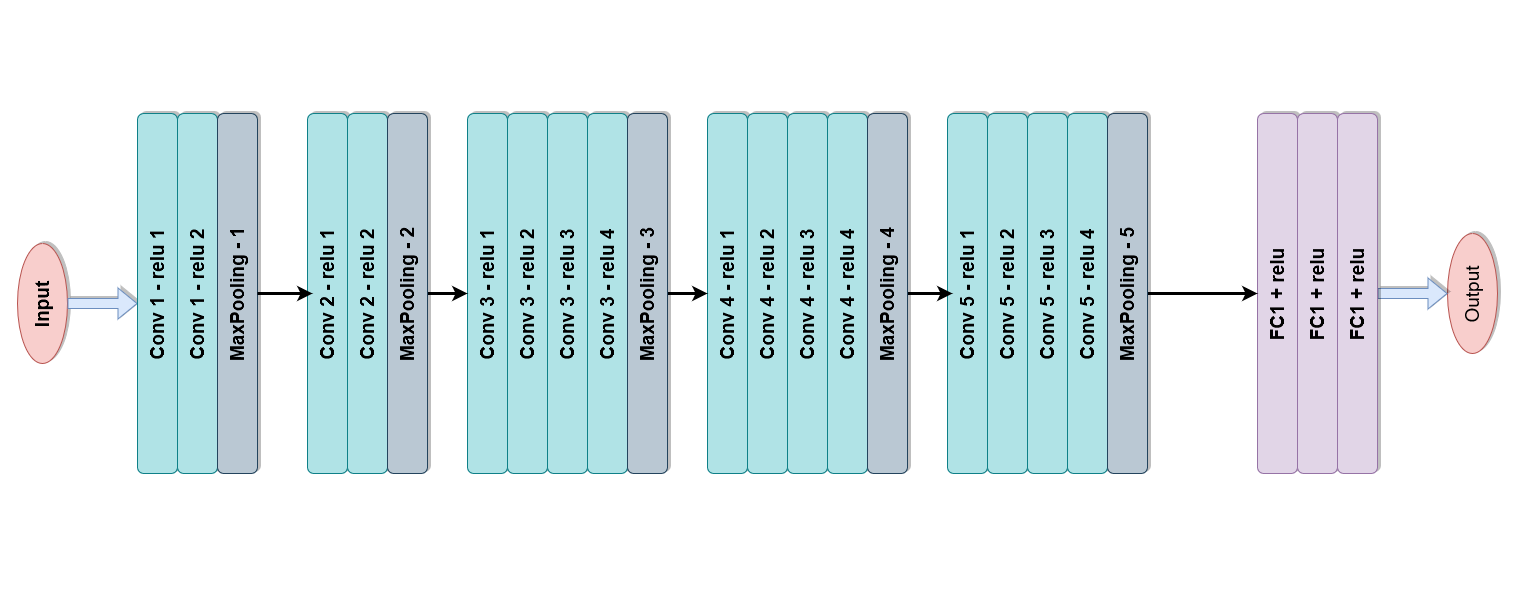}
  \caption{VGG19 Model Architecture}
  \label{fig:x VGG19 Model Architecture}
  \end{figure}

  \item \textbf{ResNet50:} ResNet50 is a deep convolutional neural network architecture that is trained on more than a million images from the ImageNet dataset. It is known for its use of residual blocks, which address the problem of vanishing gradients in deep networks. These residual blocks have shortcut connections that bypass one or more layers, allowing gradients to flow more easily and making it possible to train very deep networks without the problem of vanishing gradients. The architecture also uses batch normalization layers, which normalize the activation of the layers, making it possible to train the network more quickly and effectively. Additionally, the number of filters increases as the network gets deeper, allowing it to automatically learn more complex features. ResNet50 is widely used in many computer vision tasks and is commonly used as a feature extractor for other tasks.
  
  The ResNet50 model is used as a feature extractor in our experiment, and the output of the final convolutional layer is passed through a flattened layer which reshapes the output tensor into a 2D array. The flattened feature map is then passed through a Dropout layer with a drop rate of 0.5, which is used for regularization to prevent overfitting. The dropout layer is optional and could be removed by commenting out the line. Finally, the output of the dropout layer is passed through a dense layer with 1 unit and a sigmoid activation function, which is used to produce the final output of the network, which represents the predicted probability of the input image belonging to the target class.\cite{31}

  \newpage
  \begin{figure}[htbp]
  \centering
  \includegraphics[width=\columnwidth]{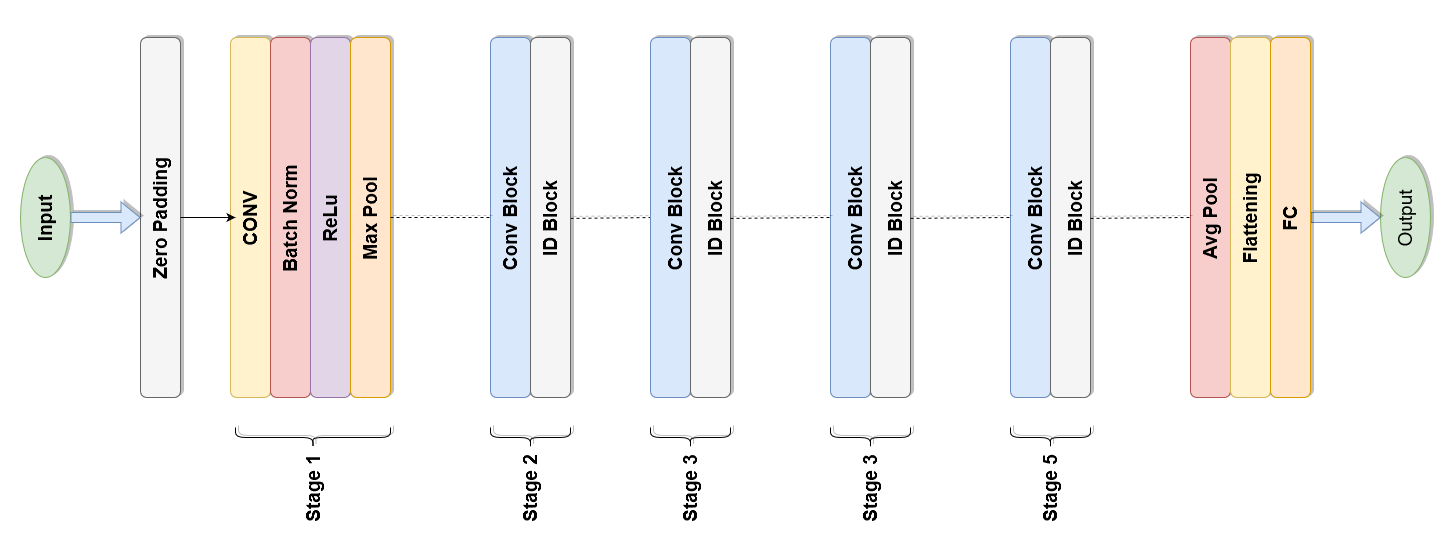}
  \caption{ResNet50 Model Architecture}
  \label{fig:x ResNet50 Model Architecture}
  \end{figure}

  \item \textbf{ResNet152:} We also used a ResNet152 architecture model. The architecture is composed of a stack of convolutional and pooling layers. The model accepts an image of shape (128, 128, 3) as input and the output of the final average pooling layer is passed through a Flatten layer which reshapes the output tensor into a 2D array. Then, there is a Dense layer with classes number of units, and softmax activation function. This dense layer produces the final output of the network which represents the predicted probability of the input image belonging to different classes. This architecture does not have a dropout layer and the model is trainable, this means that the model can be fine-tuned with new data for a different task. \cite{31}

  \begin{figure}[htbp]
  \centering
  \includegraphics[width=\columnwidth]{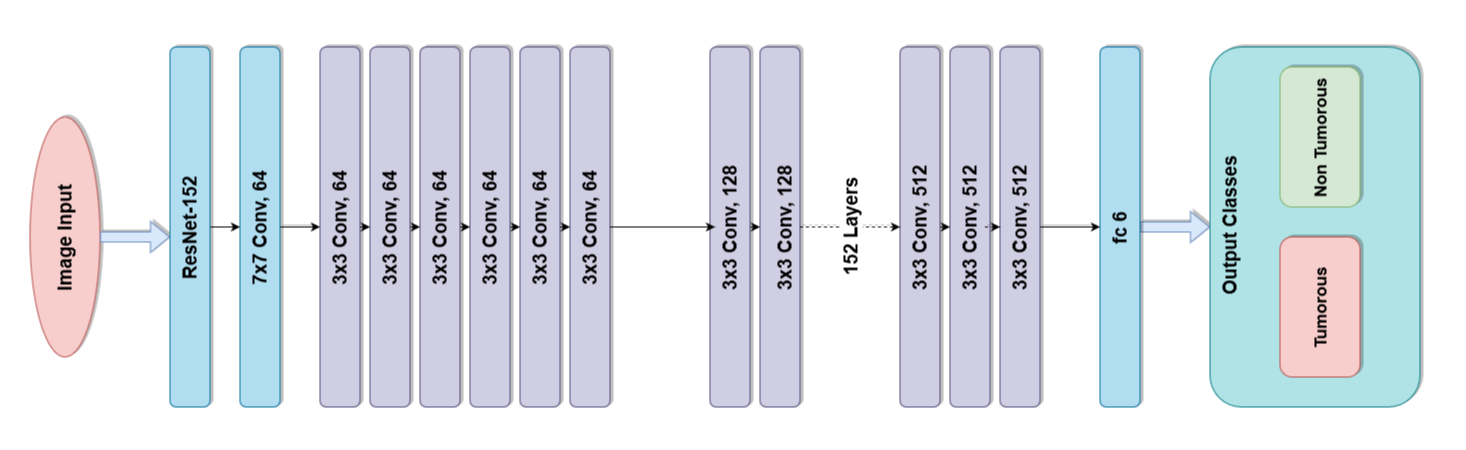}
  \caption{ResNet152 Model Architecture}
  \label{fig:x ResNet152 Model Architecture}
  \end{figure}

  \newpage
  \item \textbf{Custom CNN:} We have customized a convolutional neural network (CNN) in a similar structure to existing VGG16 and VGG19 models. It has a total of 16 convolutional layers and 3 fully connected layers. The architecture starts with an image of size (128, 128, 1) as input and predicts a binary output. The first Conv2D layer consists of 64 filters of size $3 \times 3$, followed by another Conv2D layer of 64 filters of size $3 \times 3$, and a max pooling layer with a pool size $2 \times 2$. The next two layers are similar, with 128 filters of size $3 \times 3$ and another max pooling layer. This is followed by several more pairs of DepthwiseConv2D with a different number of filters and MaxPooling2D layers that are stacked on top of each other and learn increasingly complex features from the images. After two fully connected layers with 4096 units and a dropout of 0.5 applied on each layer, the output is generated with activation “sigmoid” and Dense Layer 1 unit. Compared to VGG16, our customized CNN has a deeper architecture with more convolutional layers and is similar to that of VGG19. It also has a higher number of filters in layers than VGG16 and VGG19. Additionally, Custom CNN uses a dropout layer after each fully connected layer, something that is not present in VGG16 and VGG19. 

  \begin{figure}[htbp]
  \centering
  \hspace*{0.4cm}
  \includegraphics[width=\columnwidth]{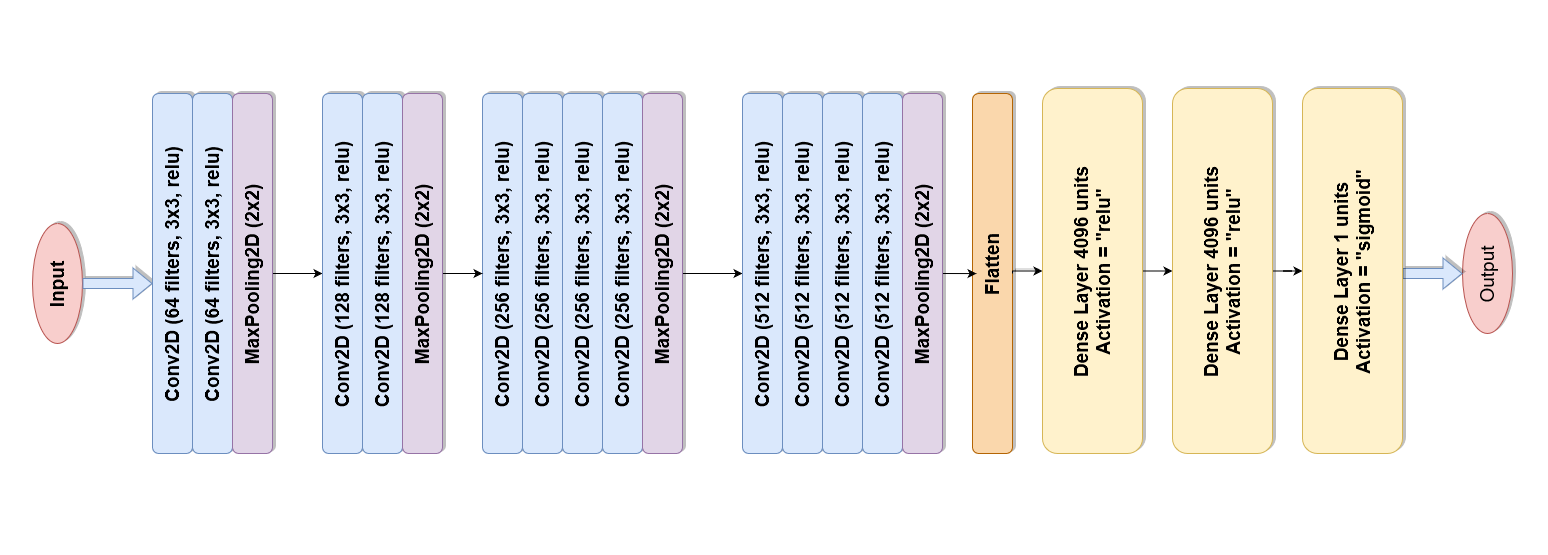}
  \caption{Custom CNN Model Architecture}
  \label{fig:x Custom CNN Model Architecture}
  \end{figure}
  
\end{enumerate}

\newpage
\section{Result Analysis}
We used homomorphic encrypted MRI data for our experiment. In our architecture, this part works as data governance. We tested simple machine-learning algorithms on both encrypted and unencrypted data. We saw on unencrypted data we were able to get the highest 97.1\% accuracy whereas, on encrypted data, we were able to get the highest 70.12\% accuracy. That shows The approach that we followed to encrypt the data is somewhat usable. 

\begin{figure}[H]
 \centering
 \includegraphics[scale=0.44]{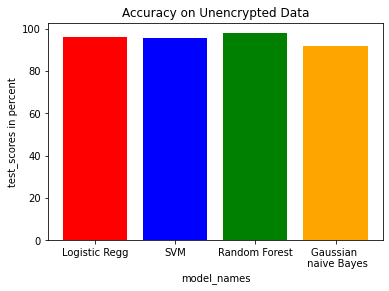}
 \includegraphics[scale=0.44]{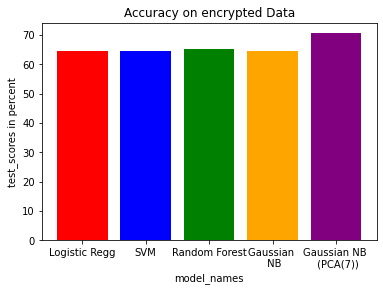}
 \caption{Performance of tested models on unencrypted and encrypted dataset}
 \label{fig:x Performance of tested models on unencrypted and encrypted dataset}
\end{figure}


For our experiment, we used four different pre-trained deep learning models - VGG16, VGG19, ResNet50, and ResNet152 - as well as a custom CNN model that we developed in-house, to classify pituitary tumor and no tumor from homomorphic encrypted MRI data during federated learning. We evaluated the performance of these models using accuracy and F1-score, as well as precision for each class. These results are from global models of federated learning which we used in our architecture.

Accuracy is the proportion of correctly classified cases out of all cases. In this study, all five models achieved accuracy greater than 50\%, indicating that they performed better than random guessing. The Custom CNN model achieved the highest accuracy of 83.31\%, followed by VGG19 with an accuracy of 78.58\%, VGG16 with an accuracy of 77.25\%, ResNet152 with an accuracy of 65.09\%, and ResNet50 with an accuracy of 61.51\%.

F1-score is a harmonic mean of precision and recall and is often used as a measure of overall model performance. Looking at the F1-score results, we can see that the Custom CNN model achieved the highest score of 83.13\%, followed by VGG19 with an F1-score of 78.31\%, VGG16 with an F1-score of 77.11\%, ResNet152 with an F1-score of 64.46\%, and ResNet50 with an F1-score of 61.45\%.


\renewcommand{\arraystretch}{1.5}
\setlength{\arrayrulewidth}{1mm}

\begin{table}[h]
\centering
\caption{Accuracy scores across different models}
\resizebox{\columnwidth}{!}{
\begin{tabular}{@{}rrrrrrrrrr@{}}
\toprule
\multicolumn{1}{c}{\multirow{2}{*}{Model}} &
  \multicolumn{1}{l}{} &
  \multicolumn{2}{c}{Accuracy Scores} &
  \multicolumn{1}{c}{} &
  \multicolumn{2}{c}{Precision} &
  \multicolumn{1}{l}{} &
  \multicolumn{2}{c}{Recall} \\ \cmidrule(lr){3-4} \cmidrule(lr){6-7} \cmidrule(l){9-10} 
\multicolumn{1}{c}{} &
  \multicolumn{1}{l}{} &
  \multicolumn{1}{c}{Accuracy} &
  \multicolumn{1}{c}{F1 - Score} &
  \multicolumn{1}{c}{} &
  \multicolumn{1}{c}{Pituitary Tumor} &
  \multicolumn{1}{c}{No Tumor} &
  \multicolumn{1}{l}{} &
  \multicolumn{1}{c}{Pituitary Tumor} &
  \multicolumn{1}{c}{No Tumor} \\ \midrule
VGG16      &  & 77.25\% & 77.11\% &  & 75.28\% & 79.22\% &  & 80.72\% & 73.49\% \\
VGG19      &  & 78.58\% & 78.31\% &  & 75.82\% & 81.33\% &  & 83.13\% & 73.49\% \\
ResNet50   &  & 61.51\% & 61.45\% &  & 62.38\% & 60.67\% &  & 57.83\% & 65.06\% \\
ResNet152  &  & 65.09\% & 64.46\% &  & 68.18\% & 62.00\% &  & 54.22\% & 74.70\% \\
Custom CNN &  & 83.31\% & 83.13\% &  & 85.71\% & 80.90\% &  & 79.52\% & 86.75\% \\ \bottomrule
\end{tabular}
}

\label{table:3}
\end{table}


The VGG16 model had a precision of 75.28\% for pituitary tumors and 79.22\% for no tumors in the binary classification task and 80.72\% and 73.49\% recall percentages accordingly. This indicates that the model had a moderate ability to correctly identify true positive cases, with slightly higher precision for no tumor cases than for pituitary tumor cases.

\begin{figure}[H]
 \centering
 \includegraphics[scale=0.32]{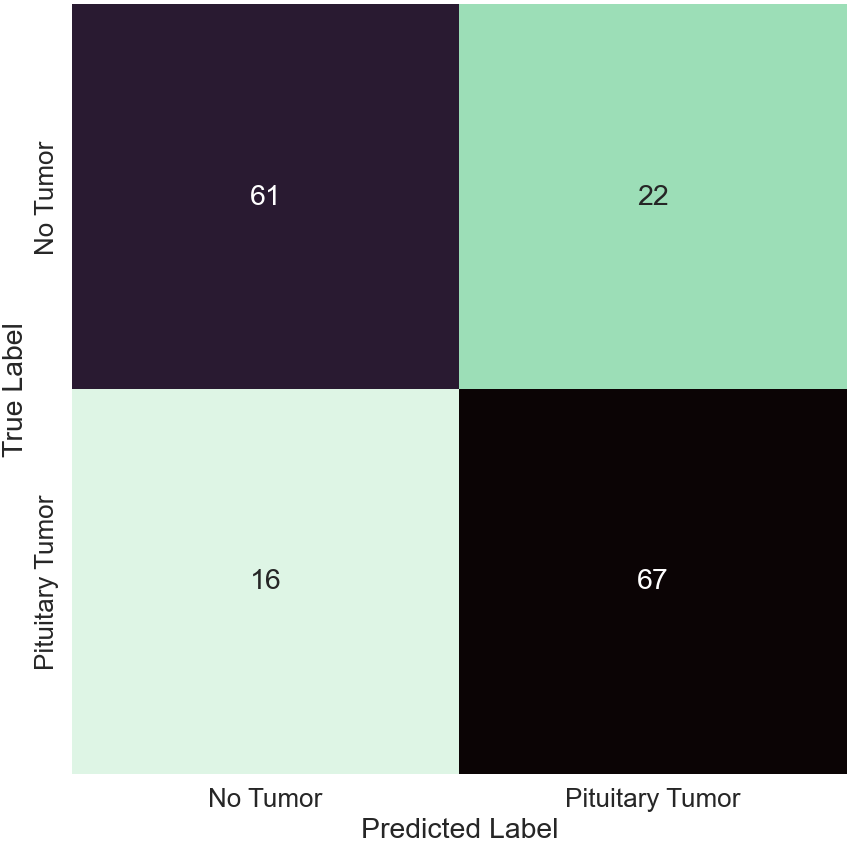}
 \caption{Confusion Matrix of VGG16}
 \label{fig:x Confusion Matrix of VGG16}
\end{figure}

\newpage

\begin{figure}[H]
 \centering
 \includegraphics[scale=0.5]{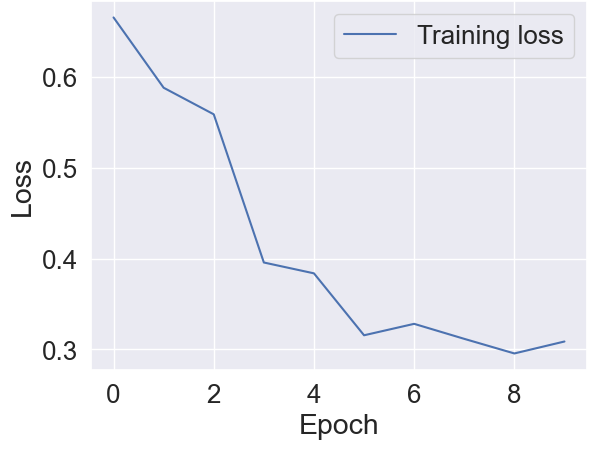}
 \caption{Training History of VGG16}
 \label{fig:x Training History of VGG16}
\end{figure}


The VGG19 model had the second-highest precision for both classes in the binary classification task. The precision for the pituitary tumor was 75.82\% and for no tumor was 81.33\%. The recall percentage is 83.13\% and 73.49\% for the classes. This suggests that the model had a moderate ability to correctly identify true positive cases, with slightly higher precision for no tumor cases than for pituitary tumor cases.

\begin{figure}[H]
 \centering
 \includegraphics[scale=0.32]{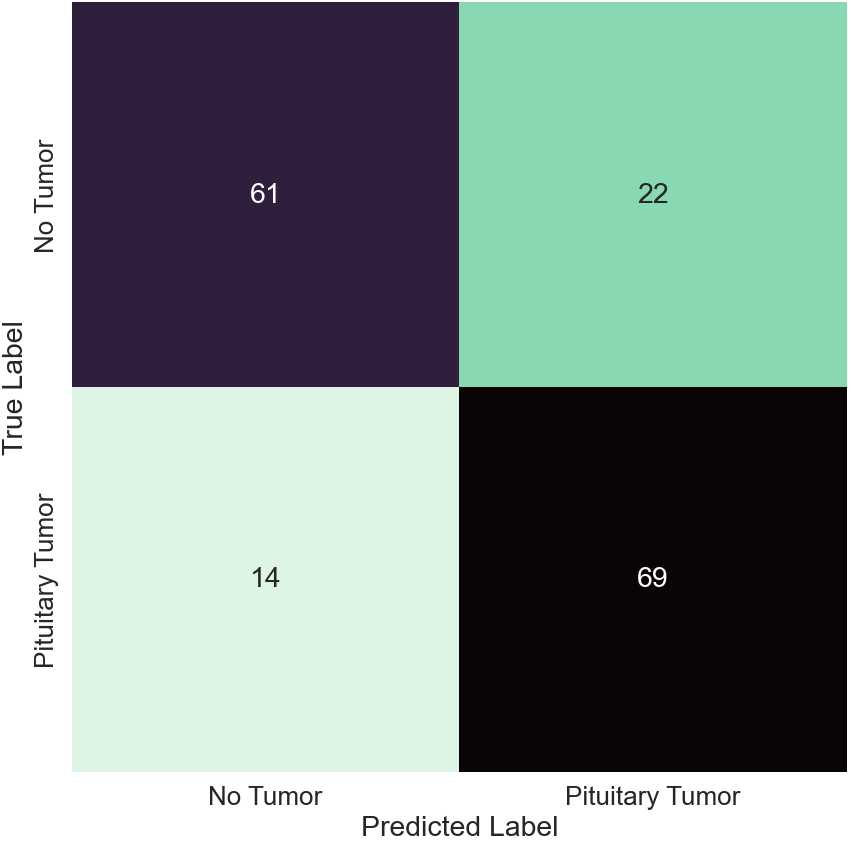}
 \caption{Confusion Matrix of VGG19}
 \label{fig:x Confusion Matrix of VGG19}
\end{figure}

\newpage

\begin{figure}[H]
 \centering
 \includegraphics[scale=0.5]{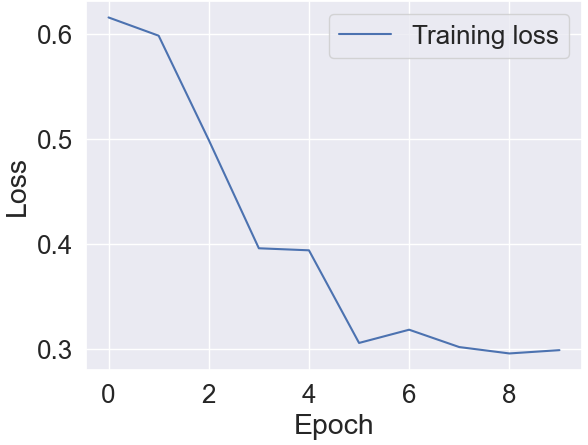}
 \caption{Training History of VGG19}
 \label{fig:x Training History of VGG19}
\end{figure}


The ResNet50 model had the lowest precision for both classes in the binary classification task. The precision for the pituitary tumor was 62.38\% and for no tumor was 60.67\% and the recall percentage was 57.83\% and 65.06\% for the classes. This indicates that the model had a lower ability to correctly identify true positive cases, with a higher number of false positives compared to the other models.

\begin{figure}[H]
 \centering
 \includegraphics[scale=0.32]{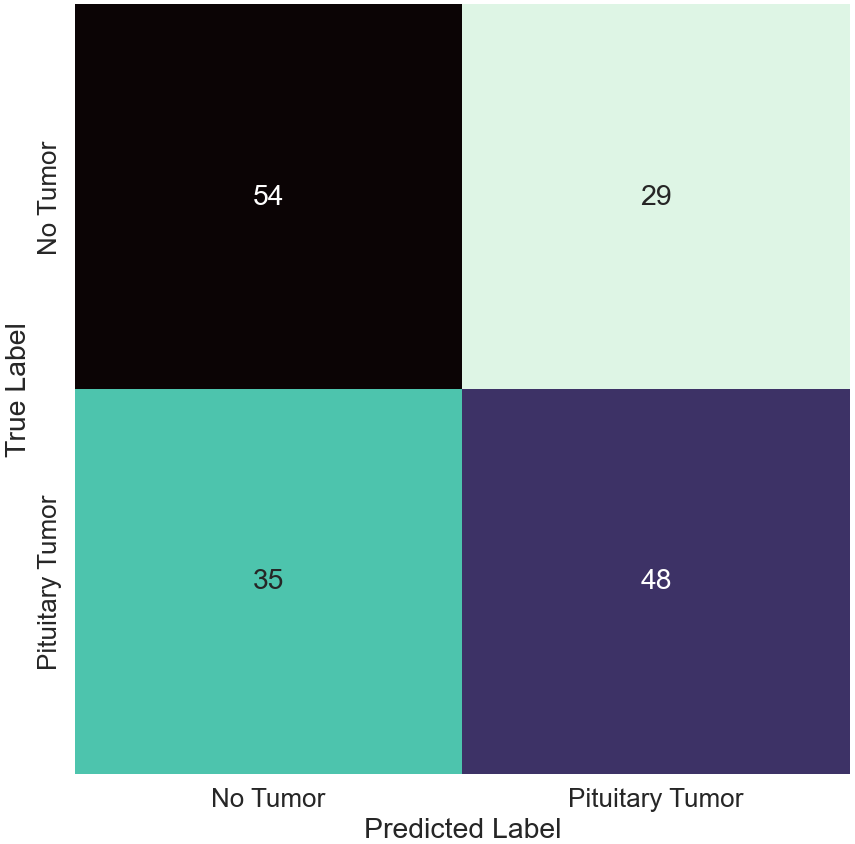}
 \caption{Confusion Matrix of ResNet50}
 \label{fig:x Confusion Matrix of ResNet50}
\end{figure}


\newpage

\begin{figure}[H]
 \centering
 \includegraphics[scale=0.5]{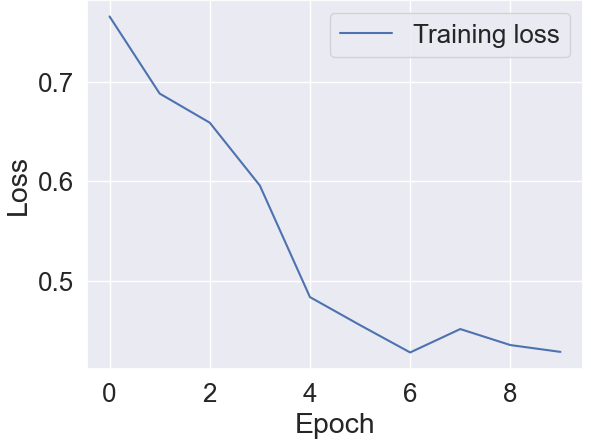}
 \caption{Training History of ResNet50}
 \label{fig:x Training History of ResNet50}
\end{figure}


The ResNet152 model had a precision of 68.18\% for pituitary tumors and 62.00\% for no tumors in the binary classification task while having 54.22\% and 74.70\% recall. This suggests that the model had a lower ability to correctly identify true positive cases, with a higher number of false positives, particularly for no tumor cases.

\begin{figure}[H]
 \centering
 \includegraphics[scale=0.32]{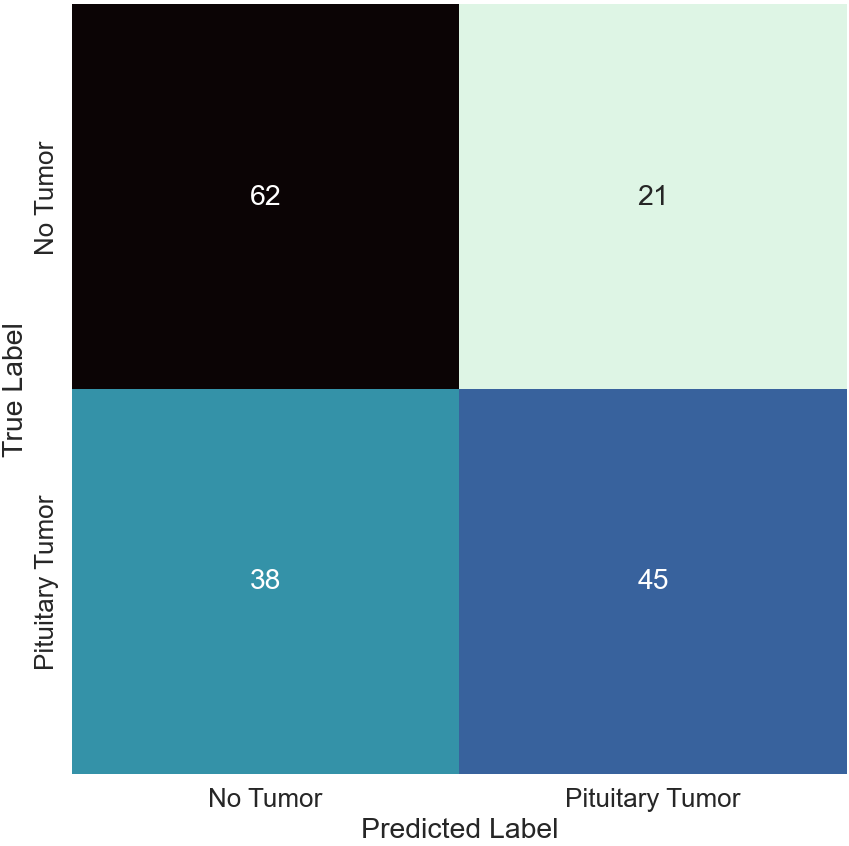}
 \caption{Confusion Matrix of ResNet152}
 \label{fig:x Confusion Matrix of ResNet152}
\end{figure}


\newpage

\begin{figure}[H]
 \centering
 \includegraphics[scale=0.5]{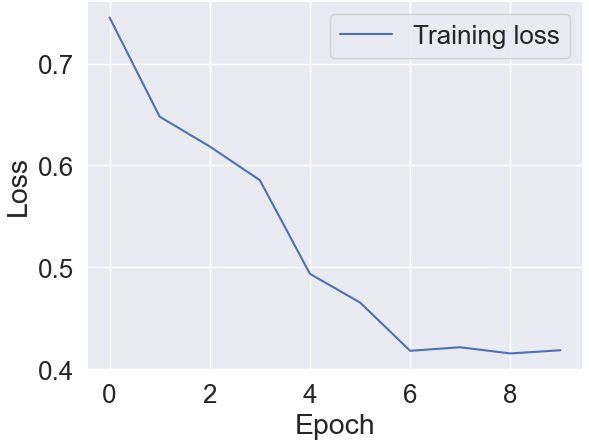}
 \caption{Training History of ResNet152}
 \label{fig:x Training History of ResNet152}
\end{figure}


The Custom CNN model achieved the highest precision for both classes in the binary classification task. The precision for the pituitary tumor was 85.71\% and for no tumor was 80.90\%. The recall percentage is 79.52\% and 86.75\% accordingly. This indicates that the model had a high ability to correctly identify true positive cases, with a relatively low number of false positives.

\begin{figure}[H]
 \centering
 \includegraphics[scale=0.32]{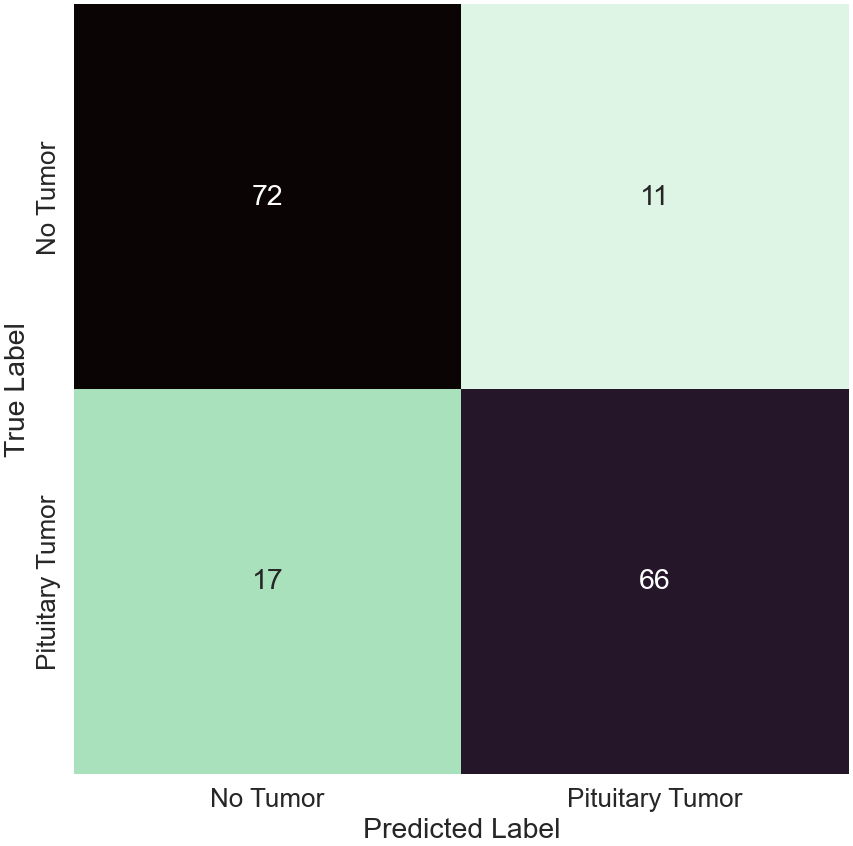}
 \caption{Confusion Matrix of Custom CNN}
 \label{fig:x Confusion Matrix of Custom CNN}
\end{figure}


\newpage

\begin{figure}[H]
 \centering
 \includegraphics[scale=0.5]{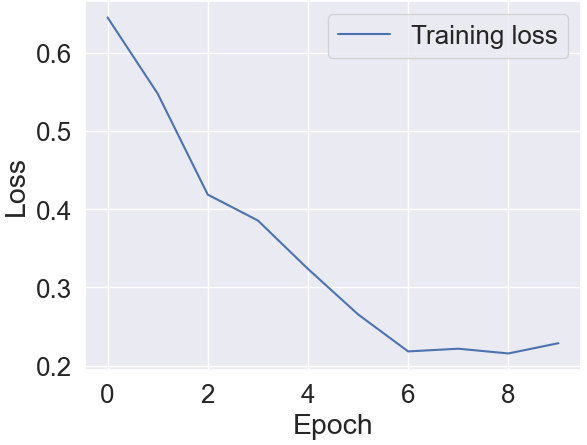}
 \caption{Training History of Custom CNN}
 \label{fig:x Training History of Custom CNN}
\end{figure}



The ROC (Receiver Operating Characteristic) curve is a powerful tool for evaluating the performance of binary classification models. It provides a graphical representation of the trade-off between true positive rate (TPR) and false positive rate (FPR) at various threshold settings, allowing you to choose an appropriate threshold based on your specific needs. A good classifier should have a ROC curve that is as close as possible to the upper left corner of the graph, indicating high TPR and low FPR. By analyzing the ROC curve, you can gain insights into the strengths and weaknesses of your model and make informed decisions about how to improve its performance. The ROC curve is widely used in various fields, including medicine, finance, and machine learning, and is an essential tool for anyone working with binary classification models. The ROC curves of our used DL models are shown below:

\newpage
\begin{figure}[H]
 \centering
 \includegraphics[scale=0.35]{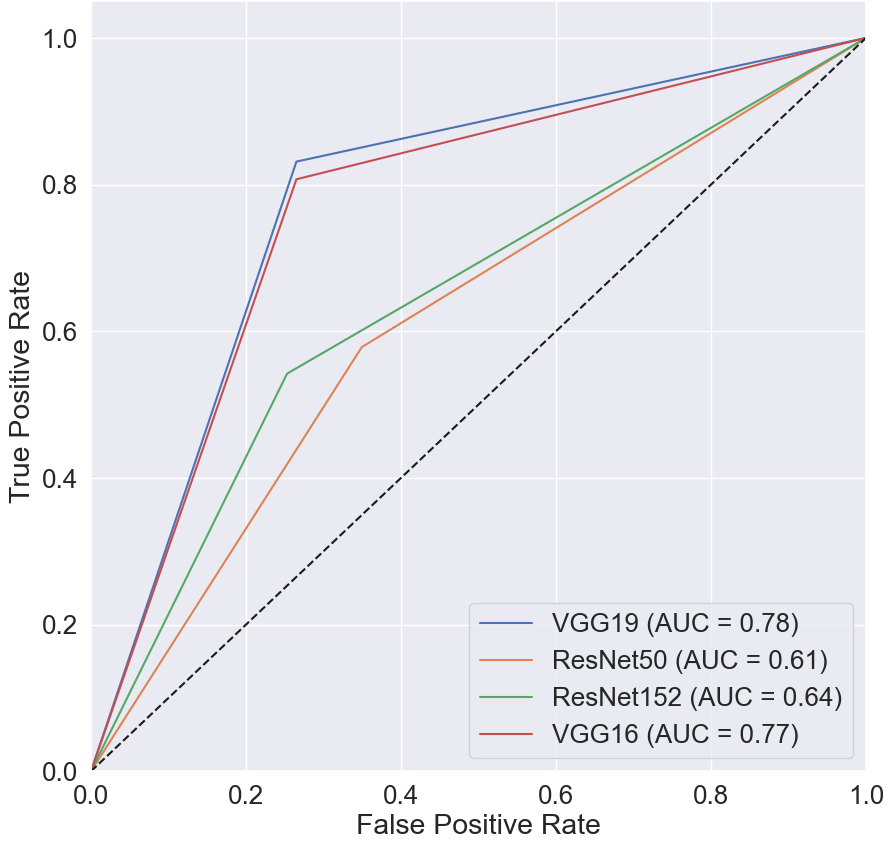}
 \caption{Receiver Operating Characteristic of Different Models}
 \label{fig:x Receiver Operating Characteristic}
\end{figure}

\begin{figure}[H]
 \centering
 \includegraphics[scale=0.35]{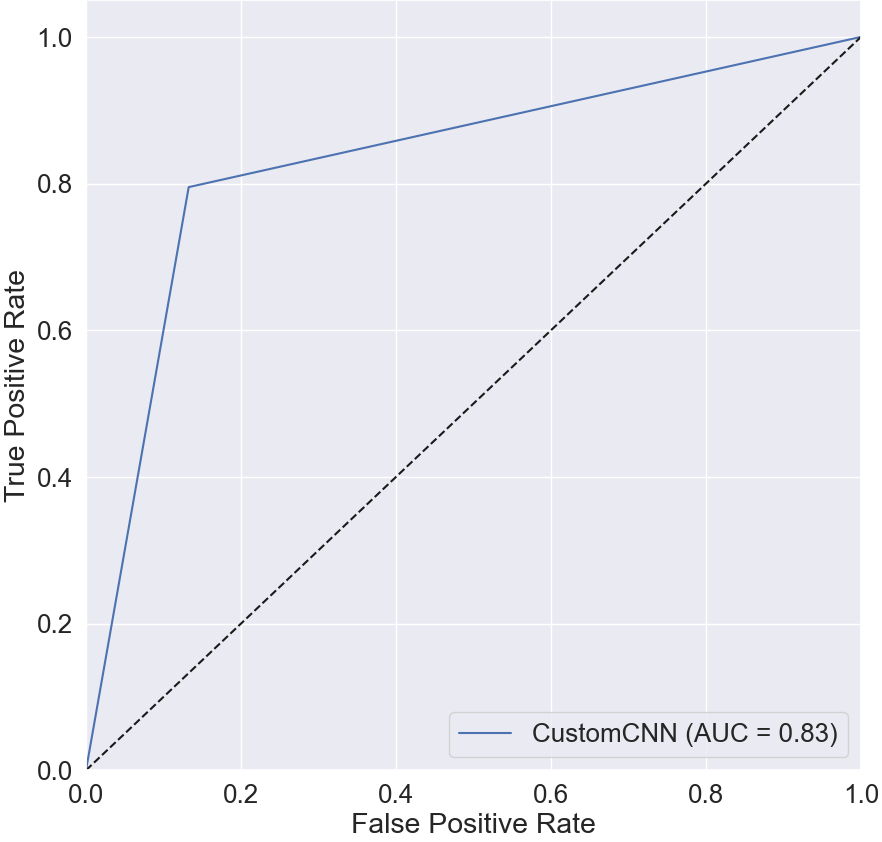}
 \caption{Receiver Operating Characteristic of Custom CNN}
 \label{fig:x Receiver Operating Characteristic of Custom CNN}
\end{figure}

\newpage

Overall, the results suggest that the Custom CNN model performed the best in classifying pituitary tumors and no tumors from homomorphic encrypted MRI data, achieving the highest accuracy, F1-score, and precision for both classes. This is particularly notable as the Custom CNN model was developed in-house, suggesting the potential for further research and development in this area. However, it is important to note that the use of partially homomorphic encryption may have influenced the results, and further research is required to confirm this hypothesis.

\section{Conclusion}

This research demonstrates an advanced data fabric architecture that enables data fusion, integration, and model parameters sharing framework to apply machine learning models without moving the data to a centralized repository.  The Partial Homomorphic EncryptionWe and Federated Learning ensured data integrity, privacy, and decentralized learning. We explored the use of pre-trained deep learning models to classify pituitary tumors from homomorphic encrypted MRI data. Our analysis showed that the VGG16 and VGG19 models outperformed the ResNet50 and ResNet152 models in terms of accuracy, precision, recall, and F1-score for both classes. We achieved overall satisfactory accuracy from VGG16 and VGG19. Out of all of these pre-trained models, our custom CNN model performed better with 83.31\% accuracy. The reason for the model to perform better is we have made the model similar to VGG16 and VGG19 while making it much less complex. Our model has a total of around 15 million parameters compared to 138 million of VGG16. Despite several significant achievements, the proposed method can be improved by using Fully Homomorphic Encryption. However, it will greatly increase the size of the images, which resulted in significant storage requirements. We have used homogeneous deep learning models for local and global model training in the federated learning approach. The heterogeneous learning models can be used for the robust data fabric architecture,

\section*{Acknowledgement}
The authors are also grateful to King Saud University, Riyadh, Saudi Arabia for funding this work through Researchers Supporting Project Number (RSP2023R18). In addition, we acknowledge the support of the PNRR project FAIR -  Future AI Research (PE00000013), Spoke 9 - Green-aware AI, under the NRRP MUR program funded by the NextGenerationEU.



\bibliographystyle{elsarticle-num} 






\end{document}